\documentclass[twocolumn]{aastex631}

\usepackage{amsmath}
\usepackage{booktabs}
\usepackage{multirow}
\usepackage{threeparttable}
\usepackage{ulem}
\usepackage{soul}
\usepackage{comment}

\begin{document}

\title{The 2015--2017 Large EVPA Rotation in OJ~287: Dominant Propagating Component in a Helical Magnetic Field with Time-Dependent Viewing Geometry}

\author[0009-0008-9101-0330]{Wanchaloem Khwammai}
\affiliation{\normalfont National Astronomical Research Institute of Thailand (Public Organization)
260 Moo 4 Donkaew, Maerim, Chiang Mai 50180}

\author[0000-0001-9799-765X]{Minchul Kam}
\affiliation{\normalfont Institute of Astronomy and Astrophysics, Academia Sinica, P.O. Box 23-141, Taipei 10617, Taiwan, R.O.C.}
\affiliation{\normalfont Department of Physics and Astronomy, Seoul National University, Gwanak-gu, Seoul 08826, Republic of Korea}

\author[0000-0002-2709-7338]{Motoki Kino}
\affiliation{\normalfont National Astronomical Observatory of Japan, 2-21-1 Osawa, Mitaka, Tokyo 181-8588, Japan}
\affiliation{\normalfont Division of Liberal Arts, Kogakuin University of Technology \& Engineering, 2665-1 Nakano-machi,  Hachioji, Tokyo 192-0015, Japan}

\author[0000-0003-0465-1559]{Sascha Trippe}
\affiliation{\normalfont Department of Physics and Astronomy, Seoul National University, Gwanak-gu, Seoul 08826, Republic of Korea}
\affiliation{\normalfont SNU Astronomy Research Center, Seoul National University, Gwanak-gu, Seoul 08826, Republic of Korea}

\author[0000-0003-0292-3645]{Hiroshi Nagai}
\affiliation{\normalfont National Astronomical Observatory of Japan, 2-21-1 Osawa, Mitaka, Tokyo 181-8588, Japan}
\affiliation{\normalfont Department of Astronomical Science, The Graduate University for Advanced Studies, SOKENDAI, 2-21-1 Osawa, Mitaka, Tokyo 181-8588, Japan}

\author[0000-0002-5158-0063]{Seiji Kameno}
\affiliation{\normalfont National Astronomical Observatory of Japan, 2-21-1 Osawa, Mitaka, Tokyo 181-8588, Japan}
\affiliation{\normalfont Joint ALMA Observatory, Alonso de Cordova 3107, Vitacura, Santiago 763-0355, Chile}

\author[0000-0001-6083-7521]{Ilje Cho}
\affiliation{\normalfont Department of Astronomy and Atmospheric Sciences, Kyungpook National University, Buk-gu, Daegu 41566, Republic of Korea}
\affiliation{\normalfont Korea Astronomy and Space Science Institute, Daejeon 34055, Republic of Korea}

\author[0000-0003-3025-9497]{Ioannis Myserlis}
\affiliation{\normalfont Institut de Radiostronomie Milim\'etrique, Avenida Divina Pastora, 7, Local 20, E-18012 Granada, Spain}

\author[0000-0001-6094-9291]{Jeffrey Hodgson}
\affiliation{\normalfont Department of Physics and Astronomy, Sejong University, 209 Neungdong-ro, Gwangjin-gu, Seoul, Republic of Korea}

\author[0000-0003-1157-4109]{Do-Young Byun}
\affiliation{\normalfont Korea Astronomy and Space Science Institute, Daedeok-daero 776, Yuseong-gu, Daejeon 34055, Republic of Korea}

\author[0000-0002-6269-594X]{Sang-Sung Lee}
\affiliation{\normalfont Korea Astronomy and Space Science Institute, Daedeok-daero 776, Yuseong-gu, Daejeon 34055, Republic of Korea}

\author[0000-0001-7003-8643]{Taehyun Jung}
\affiliation{\normalfont Korea Astronomy and Space Science Institute, Daedeok-daero 776, Yuseong-gu, Daejeon 34055, Republic of Korea}

\author[0000-0002-0393-0647]{Anne L\"ahteenm\"aki}
\affiliation{\normalfont Aalto University Mets\"ahovi Radio Observatory, Mets\"ahovintie 114, 02540 Kylm\"al\"a, Finland}
\affiliation{\normalfont Aalto University Department of Electronics and Nanoengineering, P.O. BOX 15500, FI-00076 AALTO, Finland}

\author[0000-0003-1249-6026]{Merja Tornikoski}
\affiliation{\normalfont Aalto University Mets\"ahovi Radio Observatory, Mets\"ahovintie 114, 02540 Kylm\"al\"a, Finland}

\email{trippe@snu.ac.kr}

\begin{abstract}
\noindent We present high-cadence, multi-frequency monitoring of the blazar OJ~287 using KVN (22--129~GHz), ALMA (91.5--343.5~GHz), and Mets\"ahovi (37~GHz), covering 2012--2023, together with published optical polarimetry. Within this decade-long dataset, an exceptionally large and smooth EVPA rotation is observed only during 2015--2017. The millimeter-band EVPA rotates by more than $\sim300^\circ$ over $\sim1.5$~yr, with a comparably large and more rapidly evolving rotation observed at optical wavelengths, while remaining comparatively stable at other epochs. The rotation coincides with a sequence of strong radio flares from 37 to 343~GHz whose peak amplitudes increase toward 2017 March, when the rotation ends.
Modeling the ALMA 91.5~GHz light curve yields variability Doppler factors of $\delta_{\mathrm{var}} \sim 8$--11 during the EVPA-rotation interval, with no systematic increase across the flare sequence. The absence of a monotonic change in $\delta_{\mathrm{var}}$ indicates that progressively enhanced relativistic beaming is unlikely to be the primary driver of the rising flare envelope.
The confinement of the large EVPA rotation to this interval, together with the nearly constant $\delta_{\mathrm{var}}$, indicates that the 2015--2017 event occurred when a single newly ejected disturbance temporarily dominated the polarized emission while propagating through structured inner-jet regions threaded by a helical magnetic field under modest jet-direction changes. Large EVPA rotations therefore arise intermittently as the jet orientation evolves, and are realized only when emission dominance and viewing geometry align favorably.
\end{abstract}


\section{Introduction} \label{sec:intro}

OJ~287 (0851+202) is a BL~Lac object at a redshift of $z = 0.306$ \citep{stickel89, nilsson10} and one of the most intensively observed blazars across the electromagnetic spectrum. Its optical light curve spans more than a century and exhibits prominent recurrent outbursts with a characteristic timescale of $\sim$12~yr \citep{sillanpaa88, takalo94, Sillanpaa96, sundelius96, sundelius97, valtonen08}. This long-term behavior can be explained with a supermassive black hole binary (SMBHB), in which the orbital motion of a secondary black hole perturbs the accretion flow around a primary \citep[e.g.,][]{valtonen16, valtonen23_gal}. In these models, the mass of the primary black hole is inferred to be of order $10^{10}\,M_\odot$, while the companion is assumed to be substantially less massive, of order $10^{8}\,M_\odot$ \citep[e.g.,][]{lehto96, valtonen16, valtonen23_bhmass, dey18, titarchuk23}.

However, long-term optical monitoring shows that the variability of OJ~287 is not strictly periodic, and that some major flares deviate from model predictions, most notably the absence of the expected outburst in 2021--2022 \citep{komossa23_noflare, komossa23_momo}. Motivated by such discrepancies, more recent work has refined SMBHB models by incorporating relativistic effects and detailed disk--binary interactions, suggesting that the timing of several major flares could be brought into closer agreement with observations \citep[e.g.,][]{valtonen23_mnras, zwick23}. 

Independent of the origin of the long-term optical variability, OJ~287 is an intensively studied, active blazar whose emission is dominated by a relativistic jet viewed at a small angle to the line of sight. It displays large flux variability and strong linear polarization from radio through optical and into high-energy bands, making it a prime target for time-domain studies of relativistic jets \citep{pursimo00, hovatta09, darcangelo09, agudo11, myserlis18, jormanainen25}.

During the period 2015--2017, OJ~287 exhibited an exceptionally large and smooth rotation of the electric vector position angle (EVPA) at radio and optical wavelengths, coincident with a sequence of strong radio flares \citep{myserlis18}. Large EVPA rotations or swings have been reported in several blazars at optical wavelengths \citep[e.g.,][]{larionov08, marscher08, marscher10, larionov13, morozova14, blinov15, blinov18, blinov19, kiehlmann16, liodakis20, kiehlmann21}. In OJ~287, however, such large and consistent EVPA rotations occurring simultaneously at both radio and optical wavelengths are rare \citep{kikuchi88}. Within our 2012--2023 monitoring period, this behavior was observed only once. At other epochs, the polarization variability is dominated by more irregular, short-timescale variations. 

In principle, high-resolution very long baseline interferometry (VLBI) observations are ideally suited for investigating the physical origin of such an EVPA rotation, as they can spatially resolve the innermost jet geometry and individual components that dominate at millimeter wavelengths. For example, the Event Horizon Telescope (EHT) at 230~GHz and the Global Millimeter VLBI Array (GMVA) observations combined with phased Atacama Large Millimeter Array (ALMA) at 86~GHz resolve the bent-jet geometry in the innermost region and reveal the presence of multiple polarized components and a standing recollimation shock \citep{zhao22, gomez26}. Multi-frequency polarimetric VLBI observations at 15--86~GHz further indicate that the jet is threaded by helical magnetic fields, based on the detection of transverse Faraday rotation measure (RM) gradient \citep{gomez22}. On larger scales, Radioastron space-VLBI observations at 1.4~GHz also trace the jet curvature and provide complementary constraints on the global jet geometry \citep{cho24}.

However, even intensive VLBI monitoring is limited in its ability to track rapid temporal variability within the innermost jet. For example, monthly 43~GHz VLBA monitoring revealed pronounced apparent changes in the inner-jet position angle exceeding $100^\circ$ during 2004--2006 \citep{agudo12}, but did not show any comparably large or systematic jet-direction changes during the 2015--2017 period at this resolution scale. This does not imply that jet-direction changes or kinematic variations were absent during that epoch. Rather, the core region observed with the VLBA at 43~GHz exhibits a complex and partially blended structure, which makes it difficult to isolate and reliably characterize the evolution of the innermost polarized components that are most likely responsible for the observed EVPA rotation. However, VLBI with higher angular resolutions, such as the GMVA and the EHT, are scheduled only during short observing windows—typically one week per session—conducted once per semester in the case of the GMVA and once per year for the EHT, which severely limits temporal coverage. As a result, they provide only sparse temporal snapshots and cannot continuously follow rapidly evolving events such as the 2015--2017 EVPA rotation in OJ~287.

This limitation makes high-cadence monitoring observations at millimeter and optical wavelengths essential. Regular polarization monitoring with connected interferometric arrays (e.g., ALMA), together with single-dish radio observations and optical polarimetric monitoring, provides dense temporal sampling on timescales of days to weeks, enabling the detailed tracking of EVPA evolution and its relation to flaring activity. This approach is particularly powerful because the emission is increasingly dominated by the innermost region at higher frequency. In OJ~287, \citet{myserlis18} presented long-term radio polarimetric monitoring that identified a large, smooth EVPA rotation during 2015--2017, coincident with strong flaring activities. They suggested that the EVPA rotation is attributed to a dominant polarized component propagating along a helical magnetic field within a curved jet (a “helix within helix” scenario). In this model, the radio emission, being partially optically thick, primarily traces the larger-scale jet curvature, while the optical emission, which is optically thin, can reveal additional small-scale helical motions and more complex EVPA variability.

Building on this, \citet{jormanainen25} employed significantly higher-cadence ALMA polarization monitoring, together with optical polarimetric monitoring, and confirmed the presence of a large EVPA rotation during the same period. They further showed that the optical EVPA rotation precedes and evolves more rapidly than the radio EVPA, and found that the timing of the EVPA rotation is closely associated with the emergence of a new jet component detected by the GMVA \citep{lico22}. This connection supports an interpretation in which the large EVPA rotation is driven by the evolution of a newly ejected component propagating through the inner jet. However, it is still unclear why such a large and smooth EVPA rotation is observed only during the 2015--2017 period and not at other epochs, and how jet kinematics and geometry together produce this behavior.

This paper builds directly on these efforts. We present high-cadence, multi-frequency polarimetric monitoring of OJ~287 using the ALMA observations at 91.5, 233.0, and 343.4~GHz, together with the Korean VLBI Network (KVN) single-dish observations at 22--129~GHz, Mets\"ahovi at 37~GHz and optical, spanning 2012--2023. Our dataset largely overlaps with that used by \citet{jormanainen25}, but is extended by the inclusion of the KVN single-dish polarimetry and by explicit consideration of recent high-resolution VLBI studies of the inner jet. We further expand the analysis by incorporating additional physical diagnostics, including the temporal evolution of the Faraday rotation measure, Doppler-factor estimates, and constraints based on the innermost jet geometry revealed by recent EHT \citep{gomez26} and GMVA+ALMA \citep{zhao22} observations. By combining dense time-domain polarimetry with RM and flux evolution, we aim to place new constraints on the jet geometry and on the kinematics of relativistic components propagating along the innermost jet in OJ~287.

\section{Data}

\subsection{KVN}
OJ~287 has been extensively observed as part of the Plasma-physics of Active Galactic Nuclei (PAGaN) project, a Key Science Program of the Korean VLBI Network (KVN) \citep{park18}. The KVN consists of four telescopes, KVN Yonsei (YS), KVN Ulsan (US), KVN Tamna (TN), and KVN Pyeongchang (PC). Each telescope is capable of operating at two frequencies simultaneously in dual-polarization mode \citep{lee11}. The PAGaN program is aimed at investigating active galactic nuclei (AGN) jet physics through simultaneous multi-frequency polarimetric observations. Within this program, OJ~287 was monitored in single-dish mode at 22, 43, 86, and 129~GHz from 2015 December to 2022 November, making it the most densely observed target in the PAGaN sample, with observations conducted approximately on a monthly basis. For the single-dish observations, we primarily used TN at 22/43~GHz and YS at 86/129~GHz. 


Each observing session consisted of 40-minute scans, including cross-scan pointing, sky-dipping curve measurements, and eight sets of position-switching cycles. The standard deviation of the EVPA values derived from these eight measurements was adopted as the random EVPA uncertainty. For instrumental polarization calibration, either Jupiter or the radio galaxy 3C~84 was observed during each session, based on the assumption that both are unpolarized sources (see \citealt{kam23} for more details).

\begin{table}[t!]
\raggedright
\caption{Radio observation summary}
\label{tab:obs_data}
\setlength{\tabcolsep}{7pt}
\begin{tabular}{cccc}
\hline
Instr. & Freq. & Period & $\rm N_{\mathrm{data}}$\\
 & [GHz] & & \\ 
 (1) & (2) & (3) & (4) \\
\hline
\multirow{4}{*}{KVN} & 22 & 2015 Dec. -- 2022 Nov. & 59 \\
 & 43 & 2015 Dec. -- 2022 Nov. & 59 \\
 & 86 & 2015 Dec. -- 2022 Nov. & 54 \\
 & 129 & 2015 Dec. -- 2022 Nov. & 49 \\
\hline
\multirow{3}{*}{ALMA}
    & 91.5 & 2012 Apr. -- 2022 Nov. & 2114 \\
    & 233.0 & 2014 Jan. -- 2023 Jan. & 152  \\
    & 343.5 & 2013 Jul. -- 2023 Jan. & 453  \\
\hline
Mets\"ahovi & 37 & 2015 Nov. -- 2022 June & 1059 \\
\hline
\end{tabular}
\tablecomments{Columns indicate: (1) instrument, (2) observing frequency, (3) monitoring period, and (4) number of measurements.}
\end{table}

\subsection{ALMA}
Approximately 40 bright AGN sources, evenly distributed across the sky, are observed with the ALMA 7-m Array at least every two weeks to serve as secondary flux calibrators alongside solar system objects \citep{kneissl23}. Owing to its brightness and high degree of polarization, OJ~287 has been monitored as one of these calibrator sources since 2012 April. Observations were primarily performed in Band~3 (97.5~GHz), Band~6 (233~GHz), and Band~7 (343.5~GHz), with a particularly high cadence of approximately 5–10 days in Band~3.

Data were reduced using the Analytic Matrix for ALMA Polarimetry (AMAPOLA), a set of CASA-compatible Python scripts \citep{kameno23}. For each epoch, Stokes parameters were first obtained independently for individual spectral windows (SPWs). The Stokes parameters at the representative frequency (e.g., 91.5 GHz in Band 3) were then estimated by performing a linear regression across the four SPWs. The resulting EVPA measurements were validated through comparison with independent observations from the IRAM 30-m telescope and the Submillimeter Array \citep[see][for more details]{kam25}, and have been actively used for various studies \citep{goddi21, goddi25, kam25, kam26, jormanainen25, shablovinskaia25}.

\subsection{Mets\"ahovi}
OJ~287 is one of the sources monitored by the 13.7-m radio telescope at the Aalto University Mets\"ahovi Radio Observatory as part of a long-term AGN monitoring program since the 1980s \citep{terasranta92}. For this work, we utilize 37~GHz total flux density measurements obtained with a 1~GHz dual-beam receiver centered at 36.8~GHz, covering the period from 2015 November to 2022 June to enhance the temporal sampling of the radio light curve. The observations employ an ON--ON dual-beam observing mode, with typical integration times of 1200--1800~s per data point. The detection limit is approximately 0.2~Jy under optimal conditions, and the signal-to-noise ratio cutoff is~4. A summary of all radio observational data is provided in Table~\ref{tab:obs_data}. 

\subsection{Optical data}
We utilize optical polarimetric data published by \citealt{jormanainen25}, which provide high-cadence optical polarization measurements over the 2015–2017 period. The optical polarimetric data were gathered from multiple ground-based telescopes \citep{takalo08, smith09, arnold12, piirola14, jorstad16, gupta23, mccall24}. These optical measurements include linear polarization in standard optical bands (e.g., R-band and combined R/V/g*). This enables comparisons of variability in the EVPA and fractional polarization across wavelengths.

\section{Results}\label{sec:results}

To reliably investigate the temporal evolution of the EVPA, it is essential to track its rotation as a continuous function of time. An inherent ambiguity in EVPA measurements arises because the EVPA is intrinsically defined only up to $180^\circ$. As a result, consecutive measurements may display sudden apparent jumps that do not correspond to real physical changes. For example, a smoothly varying EVPA that passes through the $180^\circ$ boundary can appear as an abrupt $\sim180^\circ$ flip in the observed time series, despite a continuous intrinsic evolution. In addition, if the observational cadence is insufficient to resolve rapid EVPA variations, a smoothly rotating EVPA may be undersampled, causing the observed EVPA evolution to underestimate the true rotation amplitude.

To solve these issues, so called $n\pi$ ambiguity, we adjust the EVPA measurements by adding or subtracting integer multiples of $180^\circ$ so that the EVPA varies smoothly with time \citep[e.g.,][]{blinov15}. The adjusted EVPA at epoch $i$ is calculated as
\begin{equation}
\chi_{\mathrm{adj},i} = \chi_i - n\pi, \quad
n = \mathrm{int}\left(\frac{\chi_i - \chi_{i-1}}{\pi}\right),
\end{equation}
where $n$ is chosen to minimize the angular difference between adjacent measurements \citep{kiehlmann16}. This procedure effectively unwraps the EVPA time series by enforcing the smallest possible step between consecutive epochs.

\begin{figure*}[t!]
    \centering
    \includegraphics[width=0.85\textwidth]{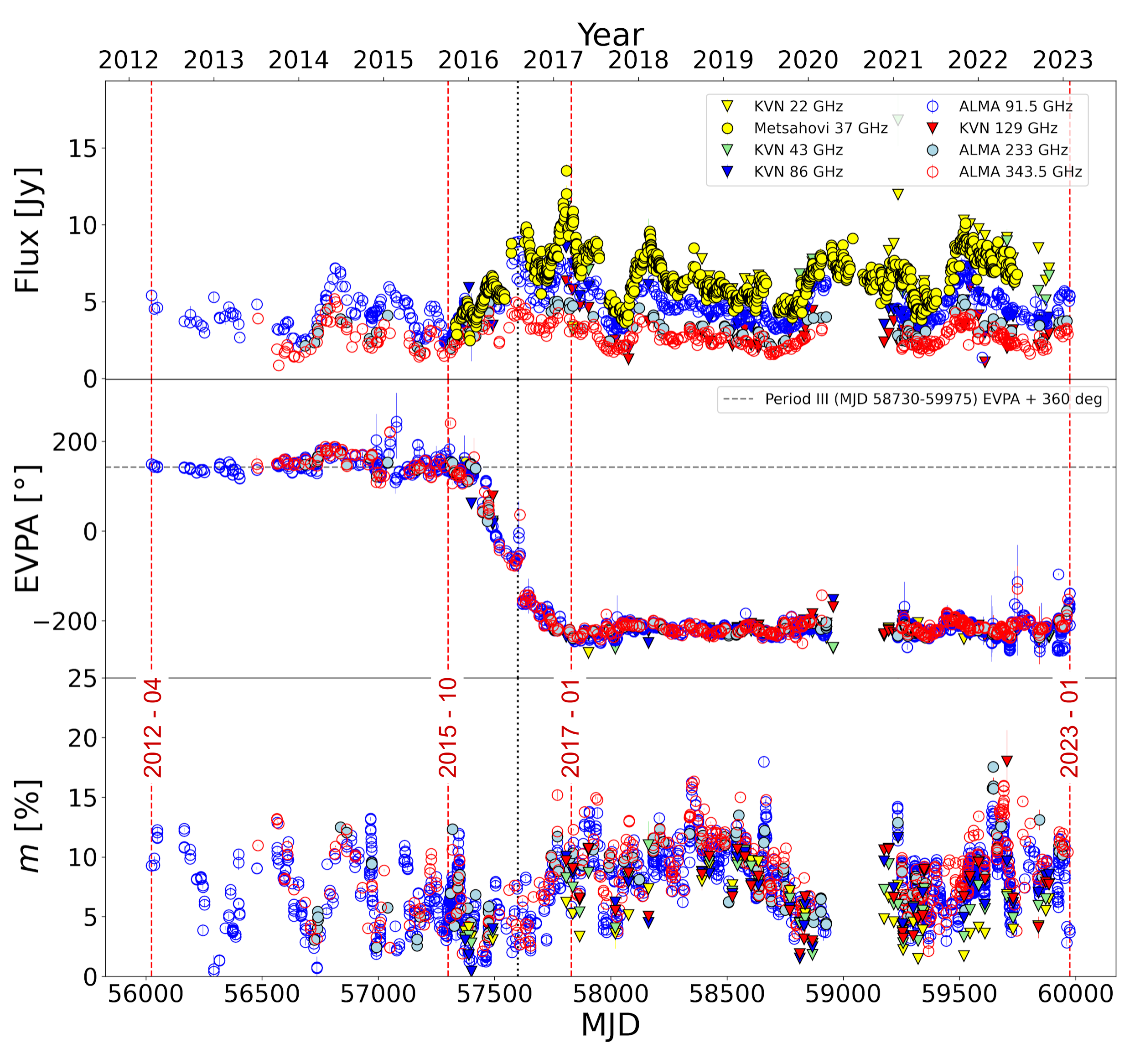}
    \caption{
    Observed radio flux density, EVPA, and polarization fraction of OJ~287 as a function of time and frequency, covering the period from 2012 April to 2023 January, hollow symbols denote data that were also published by~\cite{jormanainen25}. \textit{Top}: Total flux density obtained with KVN (22--129~GHz), ALMA (91.5--343.5~GHz), and Mets\"ahovi (37~GHz). \textit{Middle}: EVPA measurements over the same period and frequency range. \textit{Bottom}: Polarization fraction ($m$). 
    }
    \label{fig:mainplot}
\end{figure*}

This approach assumes that the intrinsic EVPA evolution is smooth on timescales comparable to or longer than the observational cadence, an assumption that is well justified for densely sampled monitoring data such as those presented here. Under this assumption, large apparent EVPA jumps are interpreted as artifacts of the $180^\circ$ ambiguity, rather than as genuine, abrupt changes in the EVPA. This assumption is widely adopted in EVPA rotation studies such as those reported in the RoboPol program \citep[e.g.,][]{blinov15, blinov18}. 

In addition to the automated correction, we performed a visual inspection of the corrected EVPA curves at all frequencies to verify the reliability of the unwrapping procedure, particularly during periods of sparse sampling or enhanced variability. This helps to identify cases in which limited cadence or large measurement uncertainties could lead to incorrect unwrapping. To further suppress spurious fluctuations, data points with EVPA uncertainties exceeding $90^\circ$ were excluded, as such measurements provide insufficient constraint on the true polarization angle and can introduce artificial rotations.

Figure~\ref{fig:mainplot} presents the resulting multi-frequency polarization monitoring of OJ~287 covering the period from 2012 April to 2023 January (MJD~56024--59975). The figure shows the temporal evolution of the total flux density, EVPA, and polarization fraction ($m$). To facilitate discussion and to highlight the different phases in the EVPA evolution, the dataset is divided into three distinct periods: Period~I (MJD~56000–57300), preceding the onset of the large EVPA rotation; Period~II (MJD~57300–57830), during which a large and smooth EVPA rotation occurs; and Period~III (MJD~57830–59975), following the EVPA rotation. Below, we discuss the polarization behavior during each period in detail.

\subsection{Period I (2012 April -- 2015 October)}

During this period, the EVPA is broadly distributed around $\sim150^\circ$ across all frequency bands, without showing a clear systematic trend. Toward the end of Period~I, in 2015, the EVPA remains predominantly within the range of approximately $100^\circ$–$150^\circ$, before the onset of the large EVPA rotation in Period~II. Occasional rapid EVPA changes are observed, most notably around MJD~57080, but such events are rare and short-lived.

Over the same period, the total flux density does not exhibit a clear long-term increasing or decreasing trend, remaining typically between $\sim1.0$ and $4.0$~Jy, with intermittent flaring activity. At higher frequencies, the variability is more pronounced: ALMA observations occasionally reach flux densities of up to $\sim7$~Jy. However, no significant correlation is found between the flux density variations and the EVPA behavior during this period.

The polarization fraction $m$ varies substantially, typically spanning the range $\sim$0.5--13\%. Notably, the polarization fraction appears to exhibit quasi-periodic variability throughout this period, without showing any clear correlation with either the total flux density or the EVPA.

\subsection{Period II (2015 October -- 2017 March)}

Period~II exhibits an exceptionally large and smooth clockwise rotation of the EVPA, with a total amplitude of approximately $400^\circ$ over a timescale of $\sim1.5$~yr. This large and smooth EVPA rotation has been reported in previous studies \citep{myserlis18, jormanainen25}. Using our KVN polarimetric data, we confirm this behavior, revealing a comparable total EVPA rotation with an average rotation rate of $\sim0.9^\circ,\mathrm{day^{-1}}$. Within our 11-yr monitoring period, such a large, smooth, and coherent EVPA rotation is observed only during this 1.5-yr interval.

Throughout most of Period~II, the polarization fraction remains comparatively low relative to Periods~I and~III. Around MJD~57600, however, the EVPA evolution is poorly constrained because of the sparse temporal sampling and small polarized flux density. The available measurements permit more than one possible EVPA configuration (Figure~\ref{fig:multi}), and the EVPA evolution through this interval therefore cannot be uniquely determined. After the gap, the EVPA resumes the same overall decreasing trend, indicating that the overall long-term rotation trend is preserved despite this local ambiguity.

Around the same epoch, the fractional polarization begins to increase across all frequencies. This change marks a clear transition from the relatively low polarization state observed earlier in Period~II and coincides with the later stages of the EVPA rotation.

Period~II is also characterized by repeating radio flares observed during our monitoring period. The total flux density shows a systematic increase at all frequencies, with a series of repeating radio flares superposed on this rising trend. As a result, the peak flux density of successive flares increases with time, producing a clear rising envelope in the light curves. This behavior is consistently observed in the ALMA (91–343~GHz), KVN (22–129~GHz), and Mets\"ahovi 37~GHz data, with all bands reaching their maximum flux density around March~2017. The peak of the strongest flare coincides with the end of the large EVPA rotation, after which the EVPA becomes relatively stable, clustering around $-200^\circ$ to $-250^\circ$, and the flux density decreases. This close temporal correspondence suggests a physical connection between the flux evolution and the EVPA rotation.

\subsection{Period III (After 2017 March)}

Following the end of the large EVPA rotation in March~2017, the EVPA becomes relatively stable across all frequencies. Throughout Period~III, the EVPA is predominantly distributed within a narrow range of approximately $-200^\circ$ to $-250^\circ$. Interestingly, this range is nearly equivalent to the EVPA distribution observed during Period~I (100–150$^\circ$), once the $360^\circ$ ambiguity is taken into account, suggesting a return to a preferred EVPA orientation after the large rotation event. These EVPA ranges are also broadly consistent with those measured in the mm-VLBI core \citep[e.g.,][]{zhao22, gomez22, gomez26, traianou25}. To examine this consistency directly, we measured the core EVPA from the 43~GHz VLBA polarization images obtained by the Boston University (BU) blazar monitoring program. The resulting VLBA core EVPAs show an overall evolution consistent with the contemporaneous KVN 43~GHz and ALMA 91.5~GHz measurements. The comparison and selected VLBA polarization maps are presented in Appendix~\ref{app:bu_evpa}. Since the mm-VLBI core EVPAs are broadly aligned with the inner jet direction, this suggests that the ordered magnetic-field component is dominated by a toroidal component.

During this period, the total flux density continues to exhibit intermittent flaring activity. However, in contrast to Period~II, these flares no longer occur on top of a systematically rising baseline. Instead, they are isolated or short-lived events occurring on relatively short timescales, without a clear long-term increasing or decreasing trend.

The polarization fraction during Period~III remains significantly higher than during Period~II, with typical values of order $\sim$10\% across frequencies. The polarization fraction shows a gradual temporal evolution: it increases during the early part of Period~III, decreases in 2019--2020, and rises again after 2021. Despite these variations, the overall polarization level remains elevated compared to the suppressed polarization state observed during the large EVPA rotation, consistent with a more ordered net polarization state in the post-rotation phase.

\subsection{Spectral index}\label{sec:spt_index}
The radio spectral index, $\alpha$, was calculated under the convention $S_{\nu} \propto \nu^{\alpha}$ by fitting a power law to the quasi-simultaneous flux densities in logarithmic space. To minimize the effects of different observing cadences among the participating facilities, we adopted the ALMA 343.5~GHz light curve, which has the longest median cadence of approximately 6~days, as the reference time series. For each 343.5~GHz epoch, measurements at 91.5~GHz and 37~GHz obtained within $\pm 3$~days were grouped together, corresponding to a total matching window of 6~days, comparable to the median cadence of the 343.5~GHz data. The flux density at each frequency was then computed as the inverse-variance weighted mean of all measurements within the window. The $\alpha$ was then determined by performing a weighted linear least-squares fit to $\log S_{\nu}$ as a function of $\log \nu$, where the weights were derived from the propagated uncertainties in logarithmic flux density. We computed the $\alpha$ using both the combined Metsähovi (37~GHz) and ALMA (91.5 and 343.5~GHz) data set, as well as an ALMA-only data set to assess the influence of cross-instrument calibration uncertainties.

\begin{figure}[t]
    \centering
    \includegraphics[width=\columnwidth]{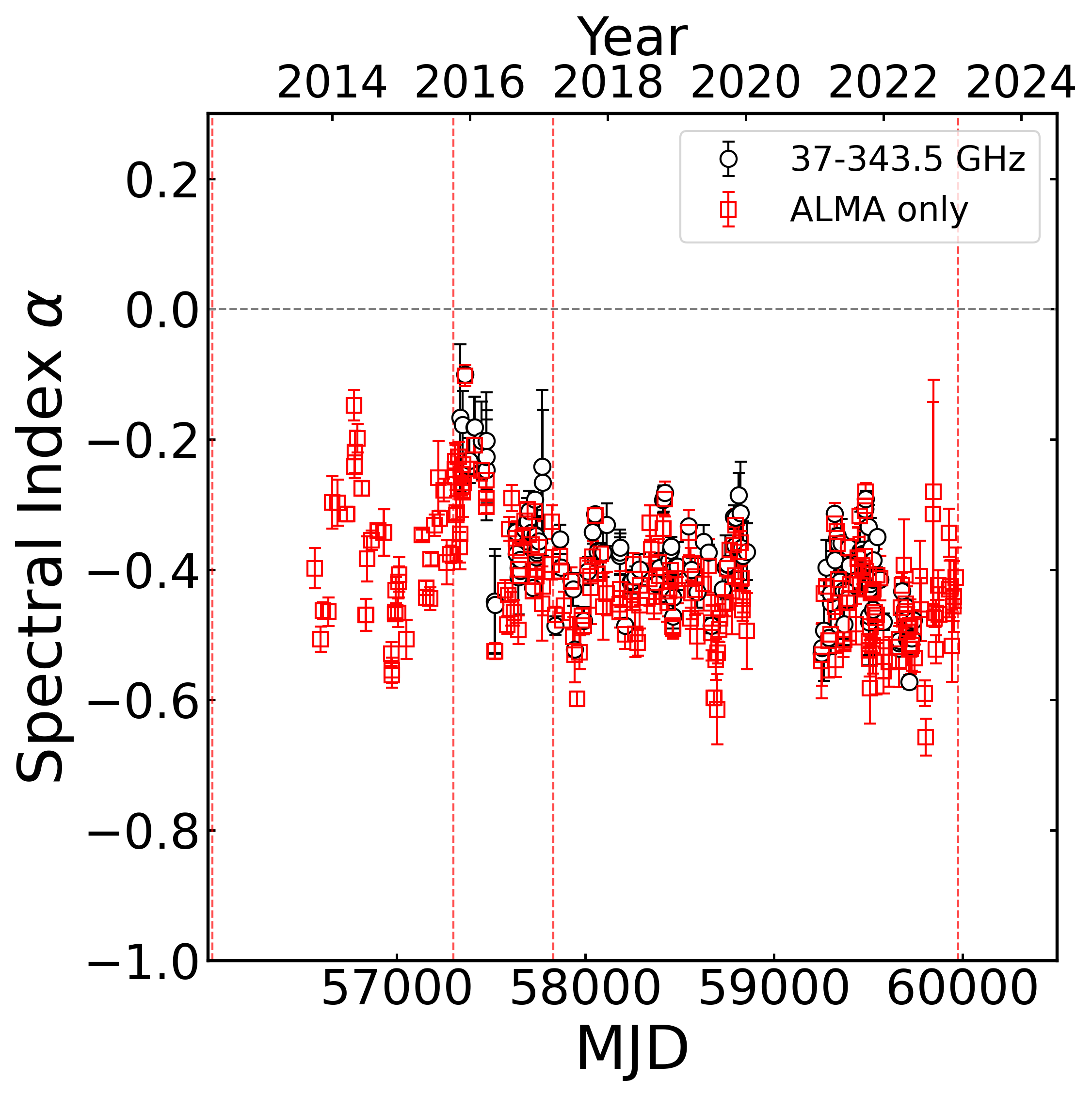}
    \caption{Temporal evolution of the radio spectral index, $\alpha$ of OJ~287. Black open circles show the spectral index derived from the quasi-simultaneous Mets\"ahovi (37~GHz) and ALMA (91.5 and 343.5~GHz) observations, while red open squares show the spectral index derived using the ALMA data alone.
}
    \label{fig:spectralindex}
\end{figure}

The fitted spectral indices remain negative throughout the observing period, ranging from approximately $-0.62$ to $-0.19$ under the convention $S_{\nu} \propto \nu^{\alpha}$ Figure~\ref{fig:spectralindex}. During Period II, $\alpha$ gradually decreases from $\sim -0.2$ near the onset of the EVPA rotation to $\sim -0.45$ toward its end, indicating a progressive steepening of the radio spectrum. After the EVPA rotation, $\alpha$ remains broadly around $\sim-0.45$ during Period III, with no return to the relatively flat spectrum observed near the onset of the rotation. The spectral indices derived from the combined Mets\"ahovi and ALMA data and from the ALMA-only data show broadly consistent temporal variation.

\subsection{Rotation measure}\label{sec:rm}

A key advantage of the KVN data is the availability of simultaneous polarimetric measurements spanning a wide frequency range from 22 to 129~GHz. This frequency coverage is crucial for bridging the gap between the 2--10~GHz measurements with the 100-m Effelsberg telescope reported by \citet{myserlis18} and the 91.5--343.4~GHz ALMA measurements presented in this work and by \citet{jormanainen25}. By connecting these frequency gaps, the KVN observations enable a more reliable determination of the $\lambda^2$ dependence of the EVPA and help separate Faraday rotation effects from intrinsic EVPA differences between frequency bands.

\begin{figure}[t]
    \centering
    \includegraphics[width=\columnwidth]{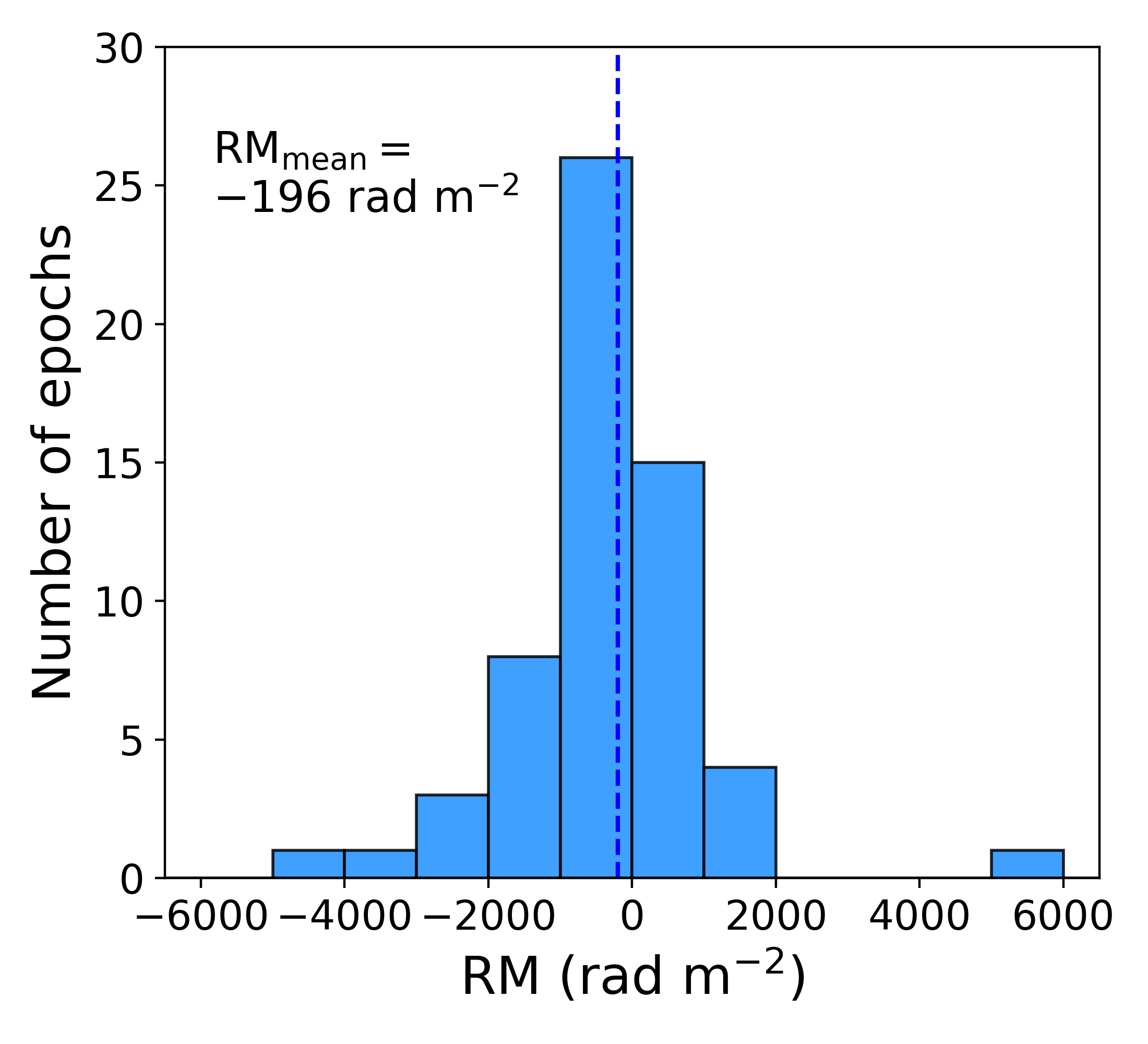}
    \caption{Histogram of the RM derived from 46 epochs of KVN observations between 2015 December and 2022 November. The vertical dashed line marks the average RM, $\mathrm{RM_{\rm mean}} = -196~\mathrm{rad~m^{-2}}$. }
    \label{fig:rm_hist}
\end{figure}

Faraday rotation occurs when intrinsically polarized synchrotron emission propagates through magnetized plasma along the line of sight. The resulting EVPA rotation follows $\rm \Delta \phi = RM\ \lambda^2$, where $\lambda$ is the observing wavelengths and RM is proportional to the line-of-sight integral of the electron density and magnetic-field strength in the foreground plasma. 

We determined the RM by performing linear least-squares fits of the EVPA as a function of squared wavelength ($\lambda^2$), using simultaneous KVN polarization observations at 22, 43, 86, and 129~GHz. The distribution of the RM values is shown in Figure~\ref{fig:rm_hist}, and the EVPA--$\lambda^2$ fits for individual epochs are presented in Appendix~\ref{app:evpa}. The average RM derived from 46 epochs obtained between 2015 December and 2022 November is $\mathrm{RM_{\rm mean}} = -196~\mathrm{rad~m^{-2}}$. Due to the limitations of the KVN data, we were unable to determine the RM during Period~I. We therefore calculated the mean RM values for Periods~II and III, obtaining $454 \pm 654$ and $-826 \pm 293~\mathrm{rad\,m^{-2}}$, respectively.

The simultaneous KVN multi-frequency observations minimize the impact of intrinsic EVPA variability on the RM estimates, which is especially critical for measuring the RM in blazars which exhibit rapid polarization changes. This enables reliable testing of a linear $\lambda^2$ relation at each epoch and allows identification of deviations that may indicate multiple polarized components or more complex Faraday-rotating media.

\section{Discussion}\label{sec:discussion}

Large rotations of the EVPA observed in blazars can arise from several physical mechanisms related to jet geometry, magnetic field structure, and the kinematics of polarized emission regions. Previous studies have shown that smooth, large EVPA rotations in blazars do not require a single unique physical origin and that multiple scenarios can reproduce the observed EVPA swings and rotations \citep[e.g.,][]{kikuchi88, marscher08, marscher10, larionov13, morozova14, blinov15, blinov18, kiehlmann16, kiehlmann21, myserlis18, lyutikov17}.

In this section, we examine the leading scenarios proposed to explain EVPA rotations, assess their applicability to the 2015--2017 event in OJ~287, and identify which interpretation best matches the full set of observational constraints.

\begin{table*}[t]
\centering
\caption{Fitted parameters of the radio flares identified in the ALMA Band~3 (91.5~GHz) data}
\label{tab:doppler}
\setlength{\tabcolsep}{6pt}
\begin{tabular}{ccccccc}
\hline
Flare & $S_0$ (Jy) & $\Delta S_{\mathrm{max}}$ (Jy) & $t_{\mathrm{peak}}$ (MJD) & $\tau_{\mathrm{rise}}$ (d) & $\tau_{\mathrm{decay}}$ (d) & $\delta_{\mathrm{var}}$ \\
\hline
1 & 1.98 & 3.12 & 57108 & 20.27 & 29.11 & 9.68 \\
2 & 2.22 & 2.98 & 57246 & 17.30 & 32.00 & 10.83 \\
3 & 2.92 & 3.06 & 57368 & 22.62 & 58.65 & 9.49 \\
4 & 2.88 & 3.82 & 57481 & 23.70 & 62.48 & 9.55 \\
5 & 2.67 & 6.56 & 57592 & 31.98 & 59.77 & 8.70 \\
6 & 2.52 & 5.33 & 57790 & 30.12 & 68.01 & 8.58 \\
7 & 2.02 & 3.22 & 57931 & 52.56 & 65.89 & 5.17 \\
\hline
\end{tabular}
\end{table*}

\subsection{Apparent EVPA rotation from turbulent polarization variability?}

One possible explanation is that an apparent EVPA rotation arises from random changes in the polarized emission from multiple turbulent cells. In the turbulent extreme multi-zone (TEMZ) model \citep{marscher14}, turbulent plasma composed of many cells with randomly oriented magnetic fields passes through a standing conical shock. Temporal changes in the relative contributions of individual cells can then occasionally produce an apparent continuous EVPA rotation without requiring a coherently rotating emission component.

The physical setting assumed in this model may be relevant to OJ~287. Stationary shock structures have been inferred from VLBI observations of OJ~287, and several features have been interpreted in terms of recollimation shocks \citep[e.g.,][]{zhao22, gomez26}. Thus, a turbulent-cell origin of EVPA variations cannot be excluded simply on the basis of the jet environment. However, \citet{marscher14} distinguished random apparent rotations in the TEMZ model from the systematic optical EVPA rotations observed in BL~Lac \citep{marscher08} and PKS~1510$-$089 \citep{marscher10}. The 2015--2017 event in OJ~287 shows a similar combination of a large and coherent EVPA rotation, a sequence of flares, and the downstream progression of the dominant VLBI emission. This close association between the EVPA rotation, flaring activity, and downstream progression of the VLBI emission favors a coherent propagating disturbance interacting with stationary jet structures over an apparent rotation arising solely from random variations among turbulent cells. We therefore consider propagation-based scenarios in the following subsections.

\subsection{A component propagating through the helical magnetic field?}

One possible mechanism involves a polarized emission component propagating along a straight jet threaded by a helical magnetic field. In this picture, as the component moves downstream, it successively samples different orientations of the twisted magnetic field. Assuming synchrotron polarization is perpendicular to the local magnetic field, the observed EVPA changes smoothly with distance along the jet, producing a systematic rotation. This scenario naturally explains large, smooth EVPA rotations when the polarized emission is dominated by a single, coherent component propagating within ordered magnetic fields \citep[e.g.,][]{marscher08, marscher10}.

Support for the presence of helical magnetic fields in AGN jets comes from multi-frequency polarimetric observations that reveal transverse RM gradients, indicating the presence of ordered helical magnetic-field structures in the innermost jet regions \citep{gomez22}. Such indications of helical magnetic fields are widely observed in nearby parsec-scale AGN jets \citep{asada02, asada08, osullivan09, broderick09,2015MNRAS.450.2441G,2018A&A...612A..67G, pasetto21, nikonov23, park26} and are consistent with leading jet-launching models in which rotation of the central black hole and/or the accretion disk naturally produces a twisted magnetic-field structure in the outflow \citep{bz77, bp82, narayan05}.

However, this scenario alone does not explain why such a large and coherent EVPA rotation is observed only during the 2015–2017 period. If the propagation of a jet component through a helical magnetic field were sufficient, similar EVPA rotations would be expected whenever such components are present. In Periods I and III, however, we observe only modest EVPA variations despite ongoing jet activity. This indicates that propagation through a helical field is not by itself sufficient to produce a large and coherent EVPA rotation. Furthermore, this scenario alone does not naturally account for the contemporaneous flux evolution, in particular the presence of repeated radio flares whose amplitudes systematically increase with time.

\begin{figure*}[t]
    \centering
    \includegraphics[width=0.75\textwidth]{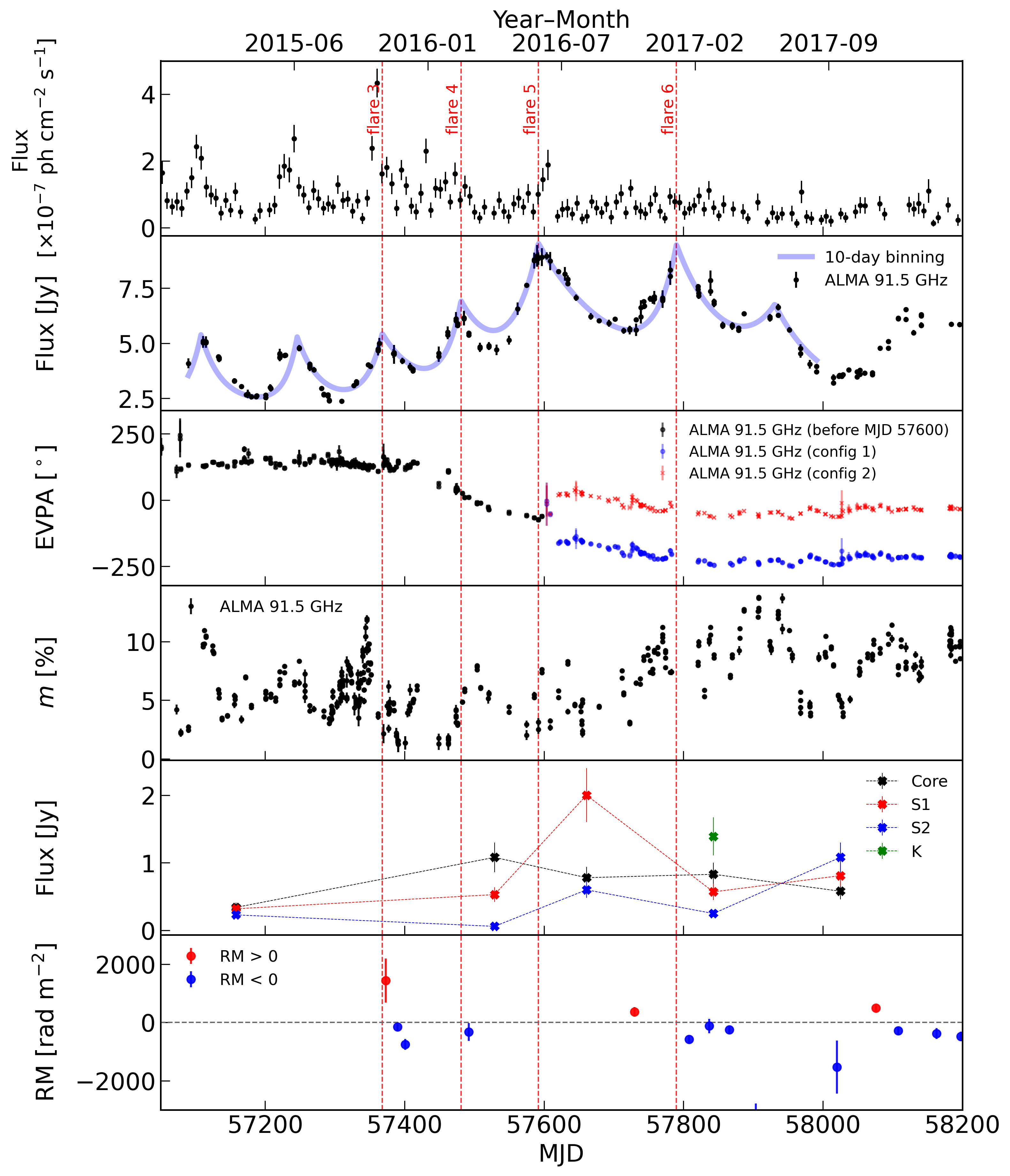}
    \caption{Multi-band light curves, EVPA and fractional polarization from 2015 March to 2018 March. The 3-day binned $\gamma$-ray flux in the top panel is taken from the Fermi LAT Light Curve Repository (\url{https://fermi.gsfc.nasa.gov/ssc/data/access/lat/LightCurveRepository/}). The blue curve in the second-top panel shows the best-fit model to the 10-day binned 91.5~GHz flux. The red vertical dashed lines indicate the flare peaks during Period II. The light curves of individual jet component shown in the bottom panel are taken from \citealt{lico22}.
    }
    \label{fig:multi}
\end{figure*}

\subsection{Helical or curved motion of a polarized component?}\label{sec:helical}

An alternative explanation attributes the EVPA rotation to motion of the polarized emission region along a curved trajectory within the jet \citep{myserlis18}. In this case, the EVPA rotation arises primarily from changes in viewing geometry as the component follows a bent or helical path. As the direction of motion and the projected orientation of the emitting region evolve with time, the observed EVPA can rotate smoothly even if the underlying magnetic field is not strongly helical. This scenario is applicable when the polarized emission is dominated by the moving component, such that its evolving orientation dominates the integrated polarized emission. 

An attractive aspect of this scenario is that it can, in principle, account for both the large EVPA rotation and the repeated radio flares whose amplitudes increase systematically with time. If the jet is curved and the component propagates along a trajectory that gradually approaches the line of sight, modest geometric changes can enhance Doppler boosting, naturally producing a rising envelope of flare amplitudes while simultaneously driving a smooth EVPA rotation. Such curved or helical inner-jet geometries are supported by recent high-resolution VLBI observations of OJ~287 with the EHT, GMVA+ALMA, and RadioAstron \citep{zhao22, cho24, traianou25, gomez26}.

Nevertheless, this scenario alone also does not explain why such a large, smooth, and coherent EVPA rotation is observed exclusively during the 2015--2017 period. Moreover, in the presence of a helical magnetic field, the resulting EVPA variability may be expected to be more complex than what is observed, particularly if multiple polarized regions contribute simultaneously.

If the propagating jet component were moving along a trajectory with a gradually decreasing viewing angle, a progressive increase in Doppler boosting could in principle produce the systematically increasing flare amplitudes observed during Period~II. To test whether such systematic beaming changes could explain the flare evolution, we estimated the variability Doppler factor, $\delta_{\mathrm{var}}$, from the ALMA Band~3 (91.5~GHz) light curve using the exponential flare decomposition method \citep{valtaoja99, hovatta09}. This approach provides a relative measure of Doppler boosting based on the observed variability brightness temperature, although it remains model dependent \citep[e.g.,][]{homan21, britzen23} and is therefore used here primarily as a consistency check rather than as an absolute measurement of jet kinematics.

Individual flares are modeled as

\begin{equation}
\Delta S(t)=
\begin{cases}
\Delta S_{\mathrm{max}}\exp\!\left(\dfrac{t-t_{\mathrm{max}}}{\tau_{\mathrm{rise}}}\right), & t<t_{\mathrm{max}},\\[6pt]
\Delta S_{\mathrm{max}}\exp\!\left(-\dfrac{t-t_{\mathrm{max}}}{\tau_{\mathrm{decay}}}\right), & t>t_{\mathrm{max}},
\end{cases}
\label{eq:flare_profile}
\end{equation}

\noindent assuming rapid particle acceleration followed by slower synchrotron cooling and adiabatic expansion. The variability brightness temperature is calculated as

\begin{equation}
T_{\mathrm{b,var}} = 1.548 \times 10^{-32}
\frac{\Delta S_{\mathrm{max}} d_L^2}
{\nu^2 \tau_{\mathrm{rise}}^2 (1+z)}
\end{equation}

\noindent and the variability Doppler factor is obtained via

\begin{equation}
\delta_{\mathrm{var}} =
\left(\frac{T_{\mathrm{b,var}}}{T_{\mathrm{b,int}}}\right)^{1/3},
\end{equation}

\noindent where we adopt an intrinsic brightness temperature of
$T_{\mathrm{b,int}} = 10^{11}\ \mathrm{K}$ \citep{kellermann04}.

Applying this analysis to the densely sampled ALMA Band~3 light curve, we identify seven distinct flares (Figure~\ref{fig:multi}). The fitted flare parameters and resulting variability Doppler factors are summarized in Table~\ref{tab:doppler}. The derived values span $\delta_{\mathrm{var}} \sim 5$--11. For the flares occurring during the EVPA-rotation interval, we obtain $\delta_{\mathrm{var}} \sim 8$--11, which remains nearly constant throughout Period~II.

We do not find evidence for a monotonic increase in $\delta_{\mathrm{var}}$ across Period~II. This disfavors a scenario in which the progressively increasing flare amplitudes are driven primarily by sustained increases in Doppler boosting associated with systematic changes in viewing angle or bulk Lorentz factor. Instead, the flare evolution is more naturally explained if a single propagating disturbance interacts sequentially with stationary structures in the inner jet, while the overall beaming conditions remain approximately constant.


\subsection{The most likely scenario: Precessing jet with helical magnetic field }

Building on previous studies that have linked the large EVPA rotation in OJ~287 to the emergence and downstream propagation of a new jet component \citep[e.g.,][]{myserlis18, lico22, jormanainen25, traianou25}, we develop a unified interpretation that connects the polarization evolution, the repeating flares, the tentative RM-sign change, and the time-dependent jet morphology into a single physical framework. In particular, we place the full flare sequence observed during the EVPA-rotation period into a self-consistent picture in which one newly ejected component propagates from the core through multiple stationary structures embedded in a helical magnetic field, while the jet axis undergoes gradual reorientation. 

In this picture, the combination of helical magnetic-field structure, relativistic motion, and time-dependent viewing geometry jointly shapes the observed total flux and polarization, while the approximately constant variability Doppler factors argue against a monotonic increase in relativistic beaming as the primary driver of the rising flare envelope.

\begin{figure*}[t]
    \centering
    \includegraphics[width=0.9\textwidth]{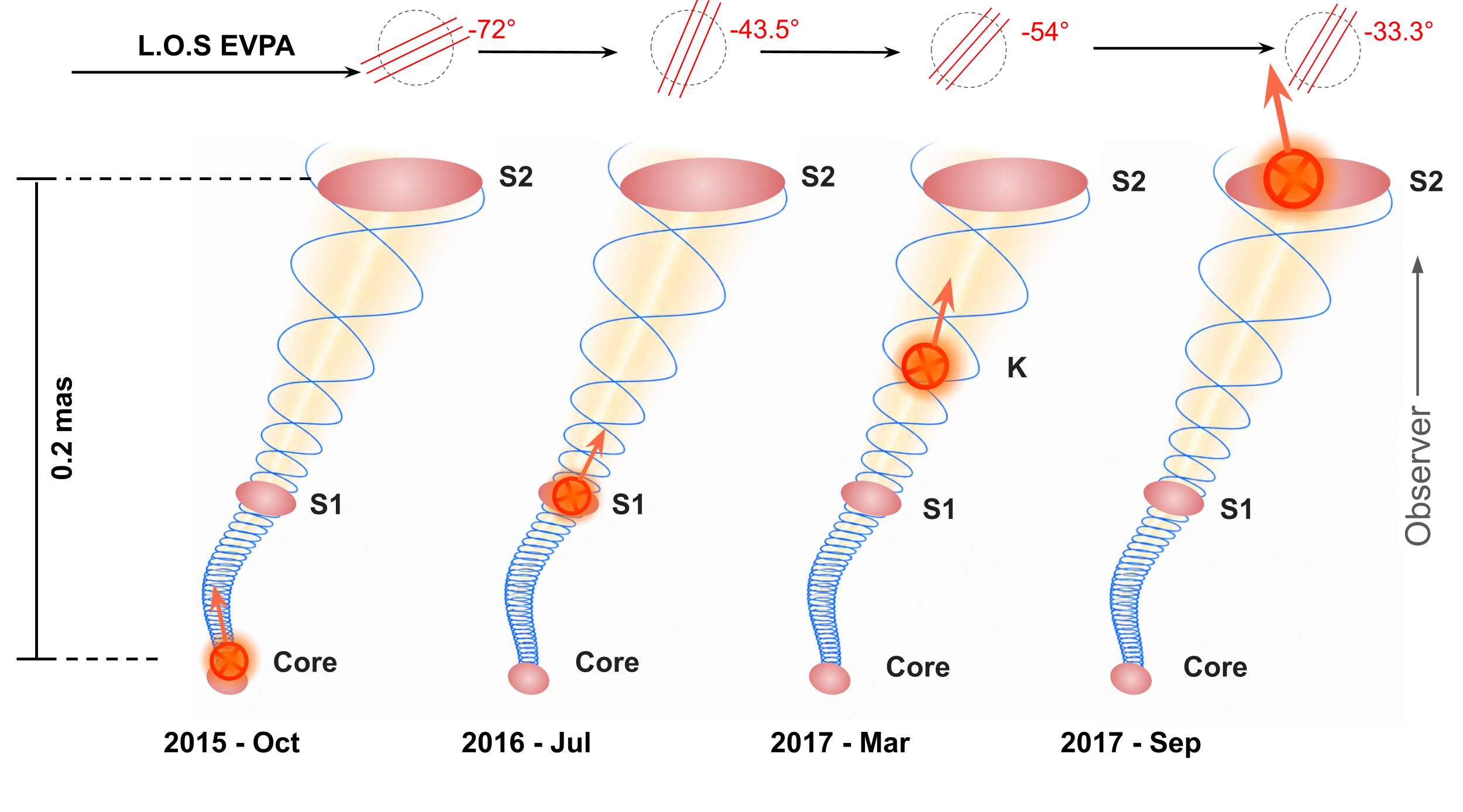}
    \caption{Schematic illustration of the inner-jet structure and evolution of OJ~287 during the 2015--2017 EVPA-rotation event. The four snapshots represent different epochs (2015 October, 2016 July, 2017 March, and 2017 September). The red circle with a cross marks a propagating disturbance moving downstream through the inner jet and interacting with quasi-stationary structures (Core, S1, and S2). Blue curved lines indicate the helical magnetic-field structure threading the inner jet. The upper panel schematically illustrates the evolution of the observed EVPA orientation as viewed along our line of sight. The labeled EVPA values are based on the 43~GHz VLBA core measurements at the nearest available epochs shown in Figure~\ref{app:fig_bu_maps}.
    }
    \label{fig:schematic}
\end{figure*}

\subsubsection{Propagation of a jet component through the innermost structures}

Within this unified framework, we interpret the large EVPA rotation and the associated flare sequence as the signature of a single newly ejected disturbance propagating through the innermost jet.

The 2015--2017 EVPA rotation begins with two major radio flares between MJD~57300 and 57500 (2015 October--2016 April; Figure~\ref{fig:multi}). In particular, the first flare (flare~3) is coincident with a gamma ray flare \citep{rakshit17}, consistent with the commonly observed association between high energy activity and the emergence of a new jet component from the core region in blazars \citep{marscher08, marscher10, jorstad16}. This interpretation is further supported by contemporaneous GMVA observations at 86~GHz. High angular resolution imaging reveals a sequence of compact features located along the jet in the order core, S1, K, and S2 (see Figure~1 in~\citet{lico22} and our Figure~\ref{fig:schematic}) 
. The core, S1, and S2 are quasi stationary features, whereas K appears as a transient brightness enhancement detected in a single epoch. The dominant emission is observed to shift progressively from the core to S1, then to K, and finally to S2 (Figure~\ref{fig:multi}), indicating downstream propagation of the emitting region. This behavior is naturally explained if a newly ejected disturbance sequentially enhances pre existing stationary structures as it propagates along the jet.

Importantly, the timing of the flaring activity seen in the GMVA components broadly matches the ALMA flare sequence during the large EVPA rotation. The early core brightening coincides with flares~3 and~4, while enhanced emission from S1 and K corresponds to flares~5 and~6. After the EVPA rotation ends, the transient component K fades and the emission becomes dominated by S2, consistent with the weaker flare~7. This temporal correspondence supports a scenario in which a single disturbance propagates through multiple stationary features, producing the observed sequence of flares. A schematic illustration of this interpretation is shown in Figure~\ref{fig:schematic}, which summarizes the downstream propagation of the disturbance and its interaction with quasi-stationary structures in the inner jet.

The spectral-index evolution provides an additional consistency check on this propagation scenario. Near the onset of Period~II, the integrated 37--343.5~GHz spectrum is relatively flat with $\alpha \sim -0.2$ (Figure~\ref{fig:spectralindex}), and subsequently steepens to $\alpha \sim -0.45$ as the EVPA rotation progresses. The observed spectral steepening can be explained if a newly emerging, partially self-absorbed component initially contributes near the core, thereby flattening the integrated spectrum. As the disturbance propagates and expands, its synchrotron self-absorption is expected to decrease, causing its emission to become increasingly optically thin and thereby steepening the integrated spectrum. Thus, the observed transition from a relatively flat to a steeper radio spectrum during Period II is consistent with the downstream evolution of the same disturbance inferred independently from the VLBI component evolution and flare sequence.

The onset of the large EVPA rotation occurs near the beginning of Period~II, during the early flare activity, suggesting that the polarized emission rapidly becomes dominated by the newly ejected disturbance even while the total flux remains largely core dominated. As the disturbance continues to propagate downstream, its interaction with successive stationary features naturally gives rise to the sequence of radio flares observed during the EVPA rotation. This flaring activity can be understood as the time dependent response of different stationary structures along the jet to the passage of the same disturbance. Such stationary features are commonly interpreted as recollimation shocks, which are expected to form naturally in relativistic jets owing to pressure mismatches between the jet and the external medium, as theoretically expected \citep[e.g.,][]{sanders83, daly88, nawaz14, fuentes18} and observationally supported \citep{marscher08, marscher10, doi18}. Observational evidence for multiple recollimation shocks in OJ~287 has been reported previously \citep{agudo11}. Within this picture, the disturbance sequentially enhances the emission from the core, S1, the transient K region, and eventually S2, producing the observed progression of flares without requiring the formation of multiple independent moving components.

\begin{figure*}[t]
    \centering
    \includegraphics[width=\textwidth]{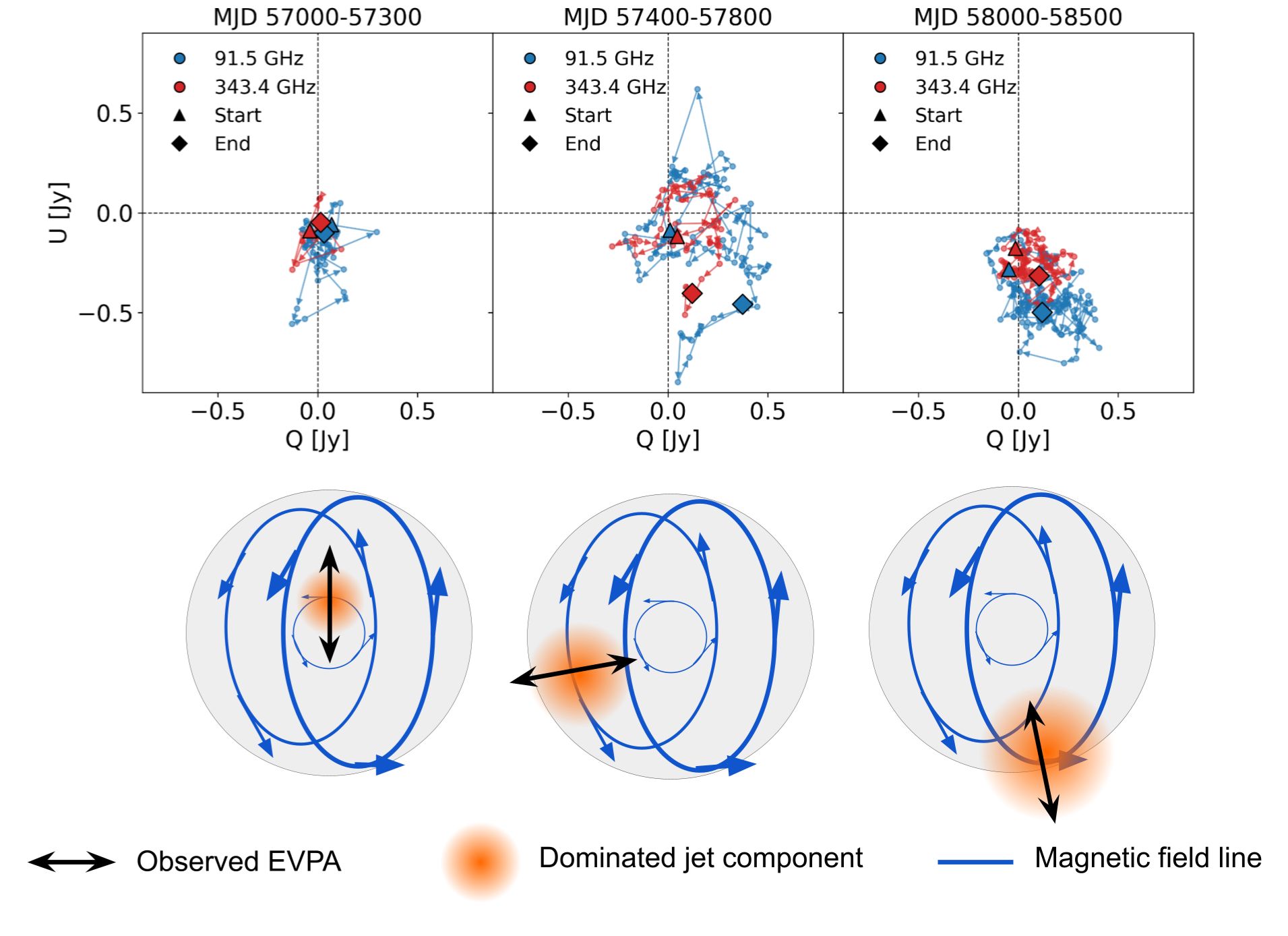}
    \caption{Stokes $Q$–$U$ distributions of OJ~287 during representative intervals of the three phases defined in Section~\ref{sec:results}. \textit{Left}: Period~I (MJD~57000–57300), showing a compact distribution offset from the origin. \textit{Middle}: Period~II (MJD~57400–57800), showing a broader distribution with rotating trajectories associated with the large EVPA rotation. \textit{Right}: Period~III (MJD~58000–58500), where the distribution re-clusters near the Period~I location. The drawings beneath the panels schematically illustrate the polarization configuration as viewed along our line of sight. Quasi-stationary polarized components dominate during Periods~I and~III, whereas a propagating disturbance temporarily dominates during Period~II. The black line segments indicate representative observed EVPA orientations.
    }
    \label{fig:qu}
\end{figure*}

\subsubsection{Viewing geometry and EVPA rotation}\label{sec:viewing_geo}

Given the propagation scenario outlined above, the large EVPA rotation can be understood as the result of a temporarily dominant emitting region. In this regime, the EVPA evolution reflects the motion of the disturbance and the projected orientation of the helical magnetic field sampled by the disturbance. Such a configuration naturally produces a large and smooth EVPA rotation when the polarized emission is predominantly governed by a single evolving component, even if weaker stationary polarized components remain present.

This condition appears to be satisfied during the 2015--2017 event, when the propagating disturbance temporarily dominates the polarized flux. Under these circumstances, geometric and magnetic-field effects manifest as a coherent EVPA evolution, whereas outside this interval the superposition of multiple polarized regions leads to comparatively stable EVPA. 

A complementary view is provided by the evolution on the Stokes $Q$--$U$ plane. Since the observed polarization vector is the vector sum of multiple polarized components, a purely single-component evolution would trace a trajectory around the origin, whereas the presence of a quasi-stationary polarized component shifts the trajectory away from $(0,0)$. Such $Q$--$U$ trajectories have been reported in association with EVPA rotations in blazars including OJ~287 and are often interpreted in terms of evolving polarization vectors and superposition of multiple components \citep[e.g.,][]{uemura10, sasada10, pasetto18, blinov19}. 

The $Q$--$U$ distribution shows a clear distinction between the three periods (Figure~\ref{fig:qu}). During Period I, the measurements form a compact cluster that is clearly offset from the origin in the $Q$--$U$ plane, indicating a persistent nonzero linear polarization vector. This suggests that the polarized emission is dominated by long-lived, quasi-stationary polarized components that define a stable polarization baseline. In Period~II, the data points expand over a much larger region of the $Q$–$U$ plane and trace rotating trajectories around this offset center. It is naturally explained if an additional, time-variable polarized component becomes dominant and rotates in the plane while superposed on the underlying stationary component. In Period~III, the distribution contracts again and re-clusters near the same offset location observed in Period~I, indicating that the transient component has faded and that the polarization is once more governed primarily by the quasi-stationary components. This evolution in the $Q$–$U$ plane supports a picture in which the 2015–2017 event corresponds to a temporary dominance of a propagating disturbance superposed on persistent polarized structures in the inner jet.

Within this picture, the low fractional polarization observed near the onset of Period~II can arise from partial cancellation between the stationary and moving polarization vectors. As the event progresses, the net polarization vector gradually returns toward the long-term EVPA direction seen outside the rotation event, reducing the degree of cancellation and naturally explaining the observed increase in fractional polarization toward the end of the rotation.

A crucial aspect of this interpretation is that large and coherent EVPA rotations are not expected to occur generically, but only when the jet geometry and viewing angle fall within a favorable configuration and when the polarized emission from the propagating component becomes comparable to or exceeds that of the stationary components in the inner jet. This geometric sensitivity is particularly relevant in OJ~287, where a large and coherent EVPA rotation is observed only during Period~II, while the EVPA remains comparatively stable outside this interval. In models involving a helical magnetic field and a slowly varying jet direction, the observed EVPA response depends sensitively on the instantaneous viewing angle \citep{lyutikov17}. Large EVPA rotations are preferentially produced when the jet direction approaches closest to the line of sight, where relativistic projection effects are strongest, because Doppler boosting can enhance the polarized flux of the moving disturbance and simultaneously place the system in the viewing-angle regime where smooth, monotonic EVPA evolution is expected. 

\begin{figure*}[ht!]
    \centering
    \includegraphics[width=\textwidth]{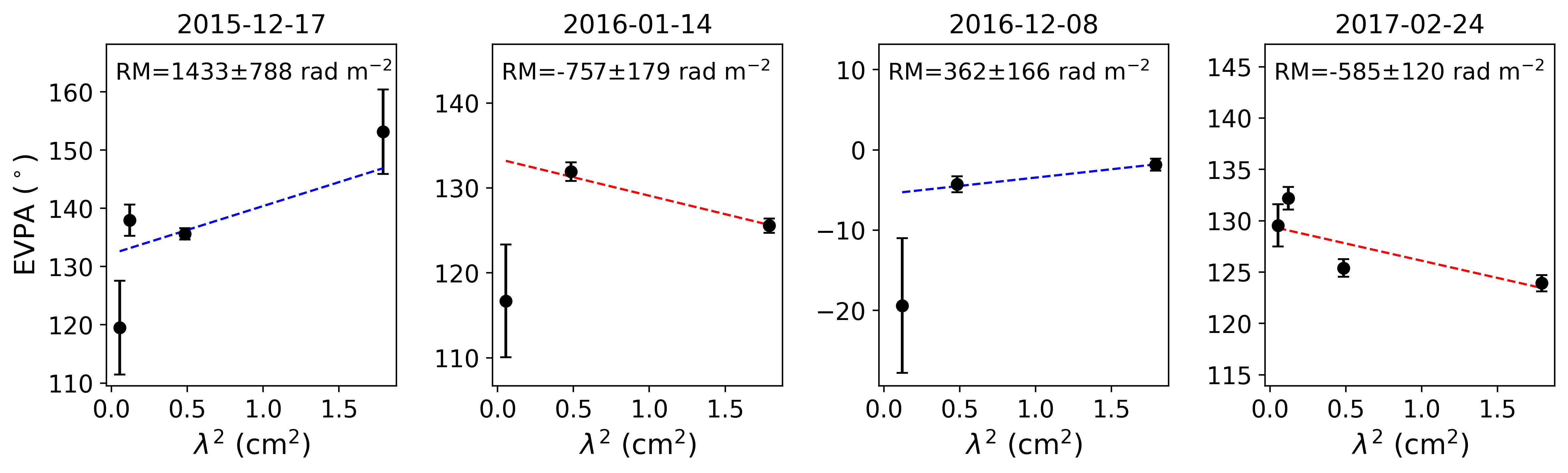}
    \caption{Evolution of the EVPA and RM of OJ 287 measured with the KVN at 22, 43, 86, and 129~GHz during the large EVPA rotation. Black data points show the observed EVPA as a function of wavelength squared, and the dashed lines show the best-fit linear EVPA--$\lambda^2$ relations. Blue and red dashed lines indicate positive and negative RM values, respectively. The epochs of 2016 January 3 and 2016 April 14 are not shown because their RM measurements have insufficient statistical significance to constrain the RM sign reliably.}
    
    \label{fig:RM}
\end{figure*}

Conversely, the absence of large EVPA rotations in Periods~I and~III can be explained by two possible factors, which may operate independently or simultaneously: (i) the polarized emission from a propagating component remains too weak relative to the quasi stationary polarized components, preventing a large and coherent EVPA rotation from being observed; and/or (ii) the jet viewing geometry does not fall within the rotation favorable geometry described by \citealt{lyutikov17}, such that even a bright propagating component would not produce a large, smooth, and monotonic EVPA rotation. These considerations indicate that both emission dominance and viewing geometry can regulate whether a large and coherent EVPA rotation is observed.

Evidence for substantial jet direction changes in OJ~287 has been reported previously, including apparent swings exceeding 100° on parsec scales \citep[e.g.,][]{agudo12, sawada-satoh13}. Curved inner jet geometries revealed by high resolution VLBI observations further support this picture, as geometric curvature enhances the sensitivity of the observed polarization angle to small changes in viewing angle \citep[e.g.,][]{zhao22, gomez22, gomez26, traianou25}. The observed combination of smooth EVPA rotation, coherent $Q$--$U$ evolution, and sequential brightening of stationary VLBI features strongly suggests that the 2015--2017 event occurred during a period of favorable viewing geometry. 

Time-dependent changes in the jet direction are common among well-sampled AGN jets \citep[e.g.,][]{lister13}, indicating that the geometric evolution invoked here need not be peculiar to OJ~287. In individual sources, such behavior has been interpreted in terms of jet precession or episodic reorientation \citep[e.g.,][]{cui23,kam24}, with possible drivers including jet instability or a misalignment between the black hole spin and the angular momentum of the accretion disk. In OJ~287, the possible presence of a SMBHB makes binary-driven jet precession or episodic reorientation a particularly relevant source-specific possibility \citep[e.g.,][]{valtonen25}.

Thus, the exceptional nature of Period~II may lie not in jet-orientation variability itself, but in the simultaneous realization of a favorable viewing configuration and the temporary dominance of a single propagating disturbance. Outside this period, similar jet activity may persist, but the polarized emission is dominated by multiple regions or by stationary features, resulting in comparatively stable EVPAs, as observed during Periods I and III. This interpretation provides a natural explanation for why large, smooth EVPA rotations are rare in OJ~287 and occur only under a specific combination of jet geometry and emission dominance.

\subsubsection{Consistency check: Tentative RM sign variations}
\label{sec:rm}

Panel 6 of Figure~\ref{fig:multi} shows the temporal evolution of the RM derived from the KVN observations, and Figure~\ref{fig:RM} presents selected EVPA--$\lambda^2$ fits during the large EVPA rotation. The epochs of 2016 January 3 and 2016 April 14 are not included in Figure~\ref{fig:RM} because their uncertainties do not allow the RM sign to be constrained robustly. Of the four epochs shown, the measurements on 2016 January 14 and 2017 February 24 provide firm negative RM detections at significance levels of approximately $4.2\sigma$ and $4.9\sigma$, respectively. The best-fit RMs on 2015 December 17 and 2016 December 8 are positive, although their individual significances are only approximately $2\sigma$. While these positive values do not by themselves establish an RM sign reversal, their occurrence at two separate epochs provides a tentative indication of temporal variation in the RM sign. We therefore regard this pattern as potentially informative as a qualitative consistency check on the proposed jet geometry.

For Faraday rotation associated with a jet threaded by a helical magnetic field, the line-of-sight magnetic-field component can vary systematically across the jet and may have opposite signs on opposite sides of the jet axis. Owing to relativistic aberration, a jet viewed side-on in its rest frame ($\theta'=90^\circ$) is observed at a viewing angle of approximately $\theta\simeq1/\Gamma$ in the observer frame, where $\Gamma$ is the bulk Lorentz factor \citep[see Figure~26 of][]{osullivan09}. When the transverse jet structure is unresolved, a side-on view can therefore yield nearly cancelling positive and negative RM contributions, whereas a modest change to $\theta<1/\Gamma$ or $\theta>1/\Gamma$ can make one sign dominate the measured RM. Thus, modest changes in viewing geometry can in principle produce temporal RM variations, including a change in sign, without requiring a reversal of the underlying helical magnetic field \citep[e.g.,][]{osullivan09, park18}.

This mechanism may be relevant to OJ~287 because recent EHT and GMVA+ALMA observations reveal that its innermost jet is curved rather than straight. As a propagating disturbance moves downstream along such a curved trajectory, the direction of its local motion relative to our line of sight can change. This can shift the effective viewing angle between the $\theta\Gamma<1$ and $\theta\Gamma>1$ regimes and thereby change which sign of the line-of-sight magnetic-field contribution dominates the observed RM. The tentative RM sign variations are qualitatively consistent with the idea that small changes in the effective jet direction can alter the sampled line-of-sight magnetic-field component and hence the observed Faraday-rotation signature.

\begin{figure*}[t]
\centering
\includegraphics[width=0.95\textwidth]{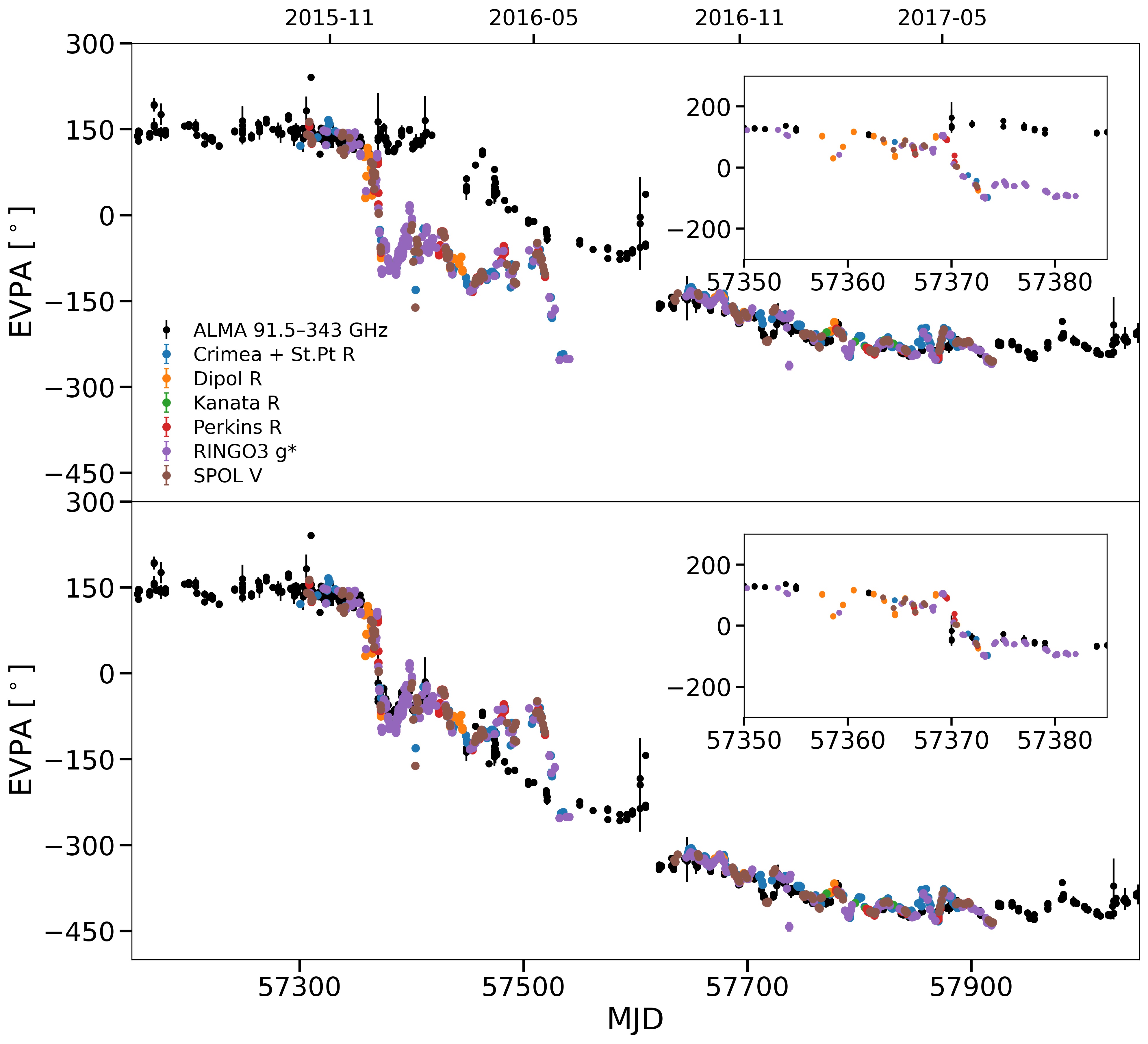}
\caption{Optical and radio EVPA evolution during the 2015–2017 event. \textit{Top}: Optical and radio EVPAs using the radio EVPA configuration obtained with the method described in Section~\ref{sec:results}. The radio EVPA values are summarized in Table~\ref{tab:polarimetry}. \textit{Bottom}: The same comparison after rotating the radio EVPA by $-180^\circ$ after MJD 57365 (2015 December). The alternative configuration brings the optical and radio EVPA curves into substantially closer agreement.}
\label{fig:optvsradio}
\end{figure*}

\subsection{Discrepancy between the optical and radio EVPA?}

During the 2015--2017 event, an apparent difference was reported between the optical and radio EVPA evolution of OJ~287 \citep{jormanainen25}. One possible origin of an optical--radio EVPA offset is Faraday rotation. However, our RM measurements of $\mathrm{RM}=-196~\mathrm{rad\ m^{-2}}$ imply a Faraday rotation of only $\sim0.1^\circ$ at 90~GHz, indicating that Faraday rotation cannot account for the observed offset (see Section~\ref{sec:rm}). Instead, the apparent offset is largely attributable to the $n\pi$ ambiguity of EVPA measurements. Rotating the radio EVPA by $-180^\circ$ after MJD~57365 brings the radio and optical EVPA curves into close agreement (Figure~\ref{fig:optvsradio}, bottom panel), largely removing the apparent discrepancy.

A residual mismatch remains around MJD~57500. Since the EVPA measurements with ALMA contain relatively long gaps compared to the optical measurements \citep{jormanainen25}, rapid EVPA variations may be undersampled by ALMA during this interval, potentially contributing to the remaining optical--radio discrepancy.

At much lower radio frequencies (e.g., 2 and 4~GHz), the EVPA does not show a comparably large rotation \citep{myserlis18}. The most plausible explanation is the lower angular resolution at these frequencies. Polarized emission from the region undergoing the EVPA rotation is blended with polarized emission from farther downstream in the jet, which contributes more prominently at lower frequencies, making the EVPA rotation less apparent. Limited temporal sampling may provide an additional contribution to the reduced detectability of rapid EVPA changes.

Although the detailed EVPA evolution around MJD~57500--57600 cannot be uniquely reconstructed because of the limited temporal sampling, the overall long-term evolution remains consistent with a large EVPA rotation associated with a temporarily dominant propagating disturbance, as discussed above.

\begin{table*}[ht]
\centering
\caption{Measurements with ALMA between MJD 57352 and 57388}
\label{tab:polarimetry}
\setlength{\tabcolsep}{6pt}
\begin{tabular}{cccccc}
\hline
Date & MJD & F [GHz] & I [Jy] & P [Jy] & $\chi$ [degree] \\
\hline
2015-11-26 & 57352.5 & 97.5  & 3.872 (0.033) & 0.246 (0.008) & -56.13 (0.3) \\
2015-11-28 & 57354.5 & 343.4 & 2.533 (0.009) & 0.144 (0.122) & -43.20 (1.5) \\
2015-11-29 & 57355.3 & 97.5  & 3.637 (0.026) & 0.258 (0.143) & -52.57 (4.0) \\
2015-11-29 & 57355.3 & 343.4 & 2.591 (0.014) & 0.169 (0.016) & -59.31 (1.1) \\
2015-12-06 & 57362.5 & 97.5  & 4.064 (0.028) & 0.156 (0.005) & -71.42 (0.4) \\
2015-12-06 & 57362.5 & 343.4 & 3.190 (0.005) & 0.099 (0.008) & -82.60 (0.6) \\
2015-12-14 & 57370.5 & 97.5  & 4.906 (0.010) & 0.052 (0.041) & -36.55 (6.2) \\
2015-12-14 & 57370.5 & 97.5  & 5.003 (0.003) & 0.066 (0.068) & -40.56 (4.6) \\
2015-12-16 & 57372.4 & 97.5  & 5.201 (0.038) & 0.105 (0.056) & -37.94 (3.6) \\
2015-12-19 & 57375.2 & 97.5  & 4.906 (0.021) & 0.175 (0.052) & -38.72 (1.8) \\
2015-12-21 & 57377.2 & 97.5  & 4.781 (0.024) & 0.263 (0.279) & -48.18 (3.3) \\
2015-12-21 & 57377.4 & 97.5  & 4.800 (0.026) & 0.153 (0.020) & -48.36 (0.4) \\
2015-12-22 & 57378.3 & 97.5  & 4.722 (0.032) & 0.215 (0.024) & -55.33 (1.1) \\
2015-12-22 & 57378.5 & 97.5  & 5.048 (0.033) & 0.207 (0.020) & -53.80 (0.8) \\
2015-12-23 & 57379.3 & 97.5  & 4.703 (0.019) & 0.207 (0.034) & -61.66 (2.1) \\
2015-12-28 & 57384.4 & 97.5  & 4.086 (0.020) & 0.203 (0.012) & -65.28 (0.9) \\
2015-12-28 & 57384.4 & 233.0 & 3.239 (0.016) & 0.186 (0.007) & -69.48 (0.5) \\
2015-12-29 & 57385.4 & 343.4 & 2.694 (0.006) & 0.136 (0.008) & -69.49 (0.8) \\
2015-12-29 & 57385.4 & 97.5  & 4.121 (0.008) & 0.194 (0.002) & -62.16 (0.1) \\
2016-01-01 & 57388.4 & 97.5  & 4.432 (0.007) & 0.092 (0.017) & -54.81 (1.7) \\
\hline  
\end{tabular}
\tablecomments{ALMA 91.5~GHz measurements shown in the small upper panel of Figure~\ref{fig:optvsradio}.}
\end{table*}

\section{Conclusions}

We presented high-cadence, multi-frequency polarimetric monitoring of the blazar OJ~287 using KVN (22--129~GHz), ALMA (91.5--343.5~GHz), and Mets\"ahovi (37~GHz), together with published optical polarimetry, spanning 2012--2023 (MJD~56024--59975). Within this decade-long dataset, we confirm that an exceptionally large and smooth EVPA rotation exceeding $300^\circ$ occurs only during 2015--2017, coincident with a sequence of repeated radio flares whose amplitudes increase progressively with time. At other epochs the EVPA remains comparatively stable, indicating that such coherent EVPA rotations are rare and realized only intermittently.

To investigate the physical conditions that enabled the 2015--2017 event, we derived variability Doppler factors using the flux density measured with the ALMA in Band 3 (91.5~GHz) from the long-term ALMA polarimetric monitoring program reduced with the AMAPOLA, a CASA-compatible analysis framework developed to process high-cadence ALMA polarization calibrator observations. For the flares occurring during the EVPA-rotation interval we obtain $\delta_{\mathrm{var}} \sim 8$--11, with no evidence for a monotonic increase across Period~II. This disfavors an interpretation in which the rising envelope of flare amplitudes is driven primarily by progressively enhanced relativistic beaming from a steadily decreasing viewing angle or increasing bulk Lorentz factor. Instead, the flare evolution is more naturally explained if a single newly ejected disturbance propagates downstream and sequentially interacts with quasi-stationary inner-jet structures, such as standing recollimation shocks, while the overall beaming conditions remain approximately constant.

Based on contemporaneous VLBI results, we interpret the large EVPA rotation and the accompanying sequence of flares as the signature of a single propagating disturbance that temporarily dominates the polarized emission while moving through a structured inner jet. The key observational point is that the large, coherent EVPA rotation is confined to 2015--2017 and is not observed during other periods, implying that the required combination of conditions is realized only intermittently. The long-term coherence of the EVPA rotation is therefore most naturally understood as arising from the combination of a helical magnetic-field structure in the inner jet and modest time-dependent viewing-geometry variations in a curved or twisted trajectory. Consistent with this picture, our simultaneous KVN observations show tentative indications of RM-sign variations during the event, providing a qualitative consistency check on the proposed time-dependent viewing geometry.

We further note that limited temporal sampling during the onset of the event can cause a rapid EVPA evolution to appear slower in the radio band, producing an apparent offset relative to the optical curve. At much lower radio frequencies, the EVPA rotation may be less apparent because the lower angular resolution blends the rotating polarized region with downstream polarized emission, while limited temporal sampling may further hinder its detection.

Overall, our results support a unified interpretation in which the 2015--2017 large EVPA rotation in OJ~287 was realized only when a single propagating component dominated the polarized emission under a favorable geometric configuration. Large, coherent EVPA rotations therefore occur intermittently, requiring a specific combination of jet geometry, magnetic-field ordering, and emission dominance. Continued long-term, high-cadence polarimetric monitoring, combined with high-resolution VLBI snapshots, will be essential for further disentangling the interplay between jet kinematics, magnetic structure, and time-dependent viewing geometry in this system.

\section*{Acknowledgments}
This research was supported by Basic Science Research Program through the National Research Foundation of Korea (NRF) funded by the Ministry of Education (RS-2024-00412117 and RS-2025-16067786). 
M.Kino is supported by the JSPS KAKENHI Grant number JP22H00157.
S. Kameno, M. Kino, and H. Nagai are supported by the JSPS KAKENHI JP21H01137.
We are grateful to the staff of the KVN who helped to operate the array and to correlate the data. 
This paper makes use of the following ALMA
data: ADS/JAO.ALMA$\#$2011.0.00001.CAL. ALMA is a partnership of ESO (representing its member states), NSF (USA) and NINS (Japan), together with NRC (Canada), NSC and ASIAA (Taiwan), and KASI (Republic of Korea), in cooperation with the Republic of Chile. The Joint ALMA Observatory is operated by ESO, AUI/NRAO and NAOJ.
The KVN and a high-performance computing cluster are facilities operated by the KASI (Korea Astronomy and Space Science Institute). The KVN observations and correlations are supported through the high-speed network connections among the KVN sites provided by the KREONET (Korea Research Environment Open NETwork), which is managed and operated by the KISTI (Korea Institute of Science and Technology Information).
This publication makes use of data obtained at Mets\"ahovi Radio Observatory, operated by Aalto University in Finland.

%

\vspace{5mm}

\bibliography{ref}{}
\bibliographystyle{aasjournal}

\appendix
\renewcommand{\thefigure}{A. \arabic{figure}}
\setcounter{figure}{0}

\section{EVPA and RM}\label{app:evpa}

\begin{figure*}[ht!]
    \centering
    \includegraphics[width=\textwidth]{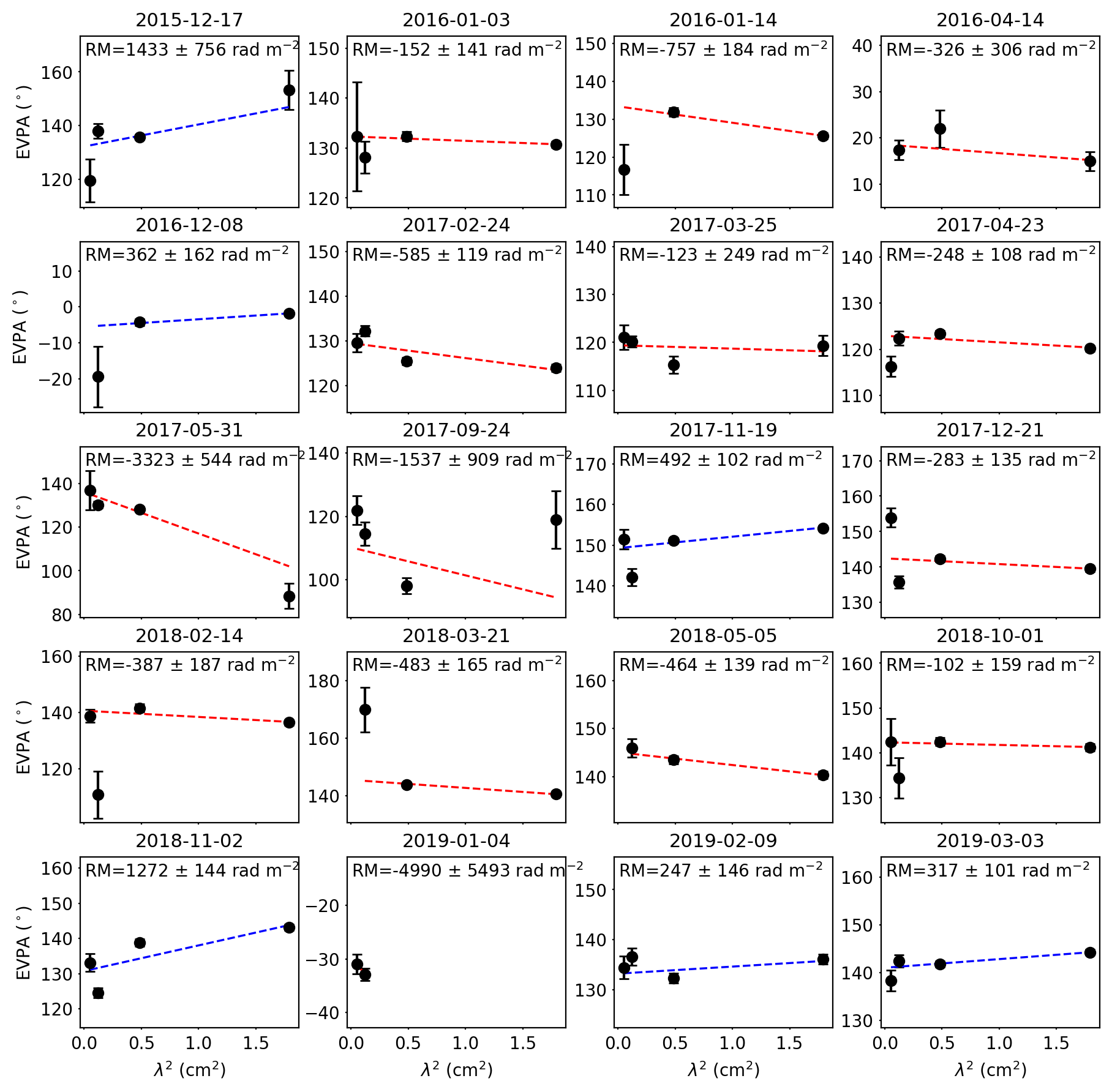}
    \caption{Evolution of the EVPA of OJ 287 measured with the KVN at 22, 43, 86, and 129~GHz from 2015 December to 2022 November. Black data points show the observed EVPA as a function of wavelength squared. The dashed lines indicate the best-fit linear relation to the EVPA. Blue and red dashed lines correspond to positive and negative RMs, respectively.}
    \label{app:fig_rm}
\end{figure*}

\addtocounter{figure}{-1} 
\begin{figure*}[ht!]
    \centering
    \includegraphics[width=\textwidth]{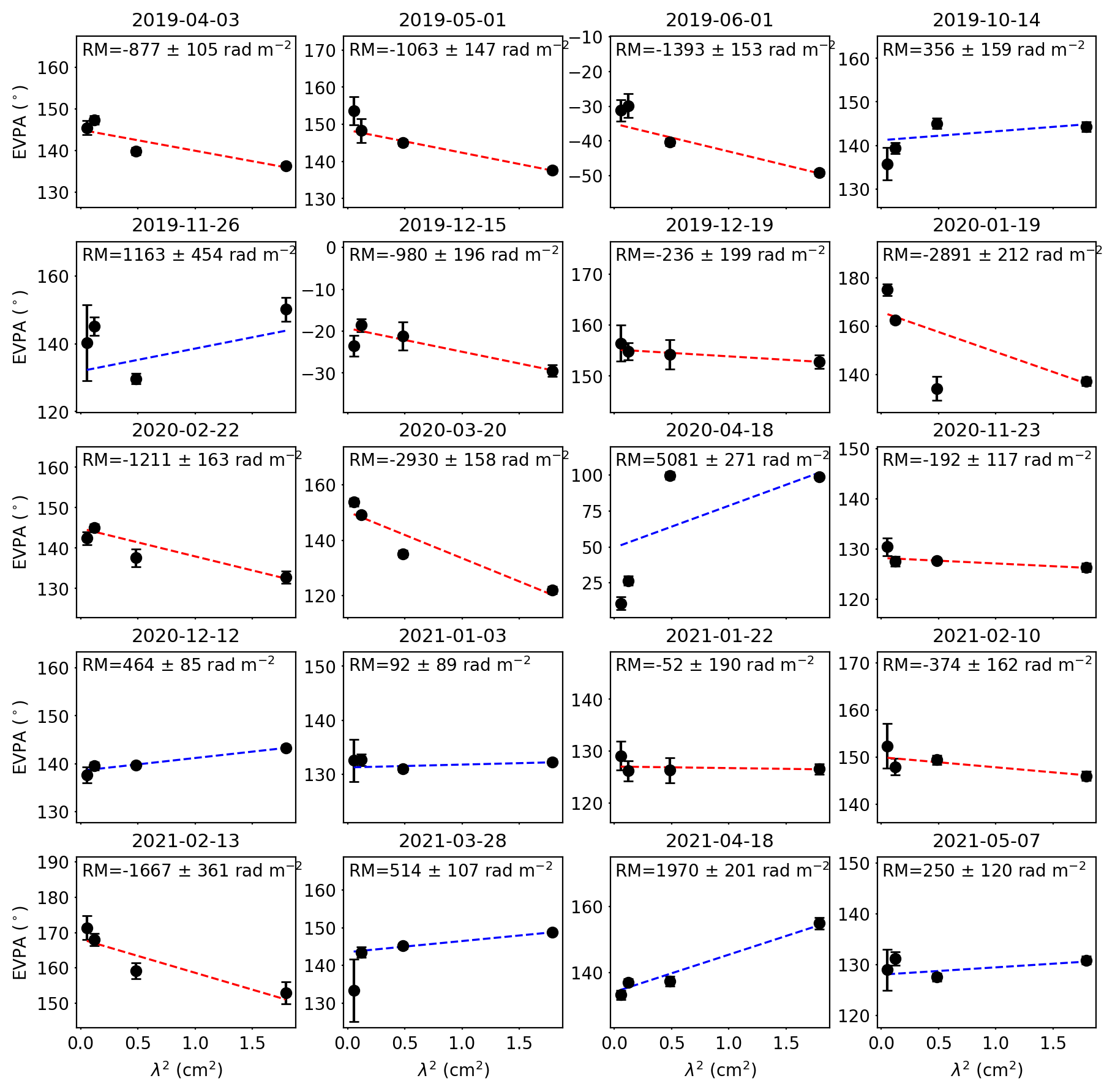}
    \caption{Continued.}
\end{figure*}

\addtocounter{figure}{-1}
\begin{figure*}[ht!]
    \centering
    \includegraphics[width=\textwidth]{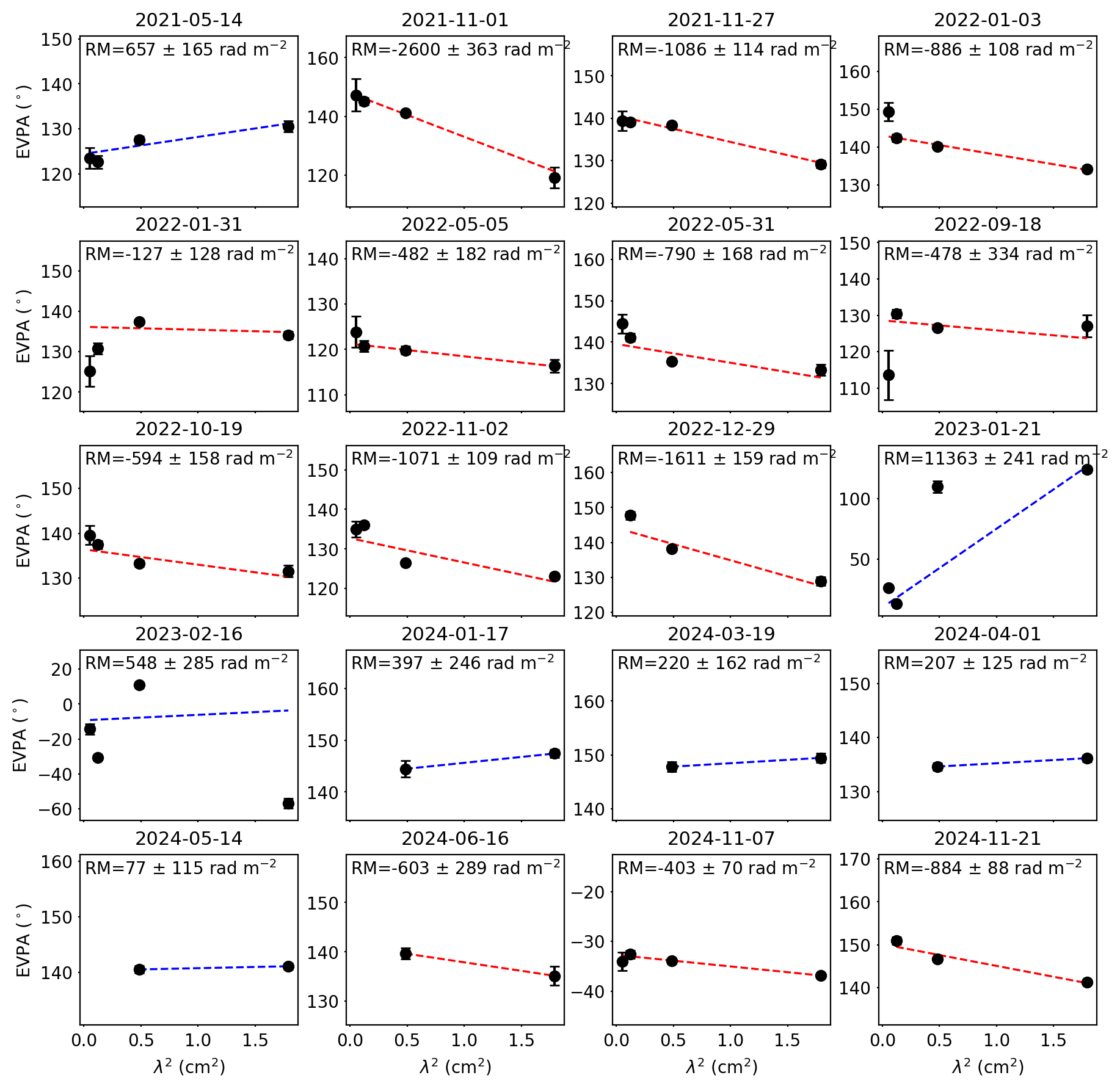}
    \caption{Continued.}
\end{figure*}

\section{Comparison with Boston University VLBA EVPA}\label{app:bu_evpa}

\renewcommand{\thefigure}{B. \arabic{figure}}
\setcounter{figure}{0}

To directly compare the unresolved EVPA measurements with the VLBI core polarization, we measured the core EVPA using 43~GHz VLBA polarization images from the Boston University (BU) blazar monitoring program \citep{jorstad16, weaver22}. We use BU 43~GHz VLBA data spanning the interval before, during, and after the large EVPA rotation. This comparison allows us to examine whether the preferred EVPA orientations observed with the KVN and ALMA before and after the rotation are also present in the VLBI core, and whether the large EVPA rotation itself is reproduced by the independent VLBA measurements.

For each available epoch, we integrated the Stokes $Q$ and $U$ emission over the core component and derived the corresponding core EVPA according to

\begin{equation}
\chi = \frac{1}{2}\operatorname{atan2}(U,Q),
\end{equation}
where the usual $180^\circ$ ambiguity of the EVPA was taken into account when comparing measurements obtained with different instruments.

\begin{figure*}[ht!]
\centering
\includegraphics[width=\textwidth]{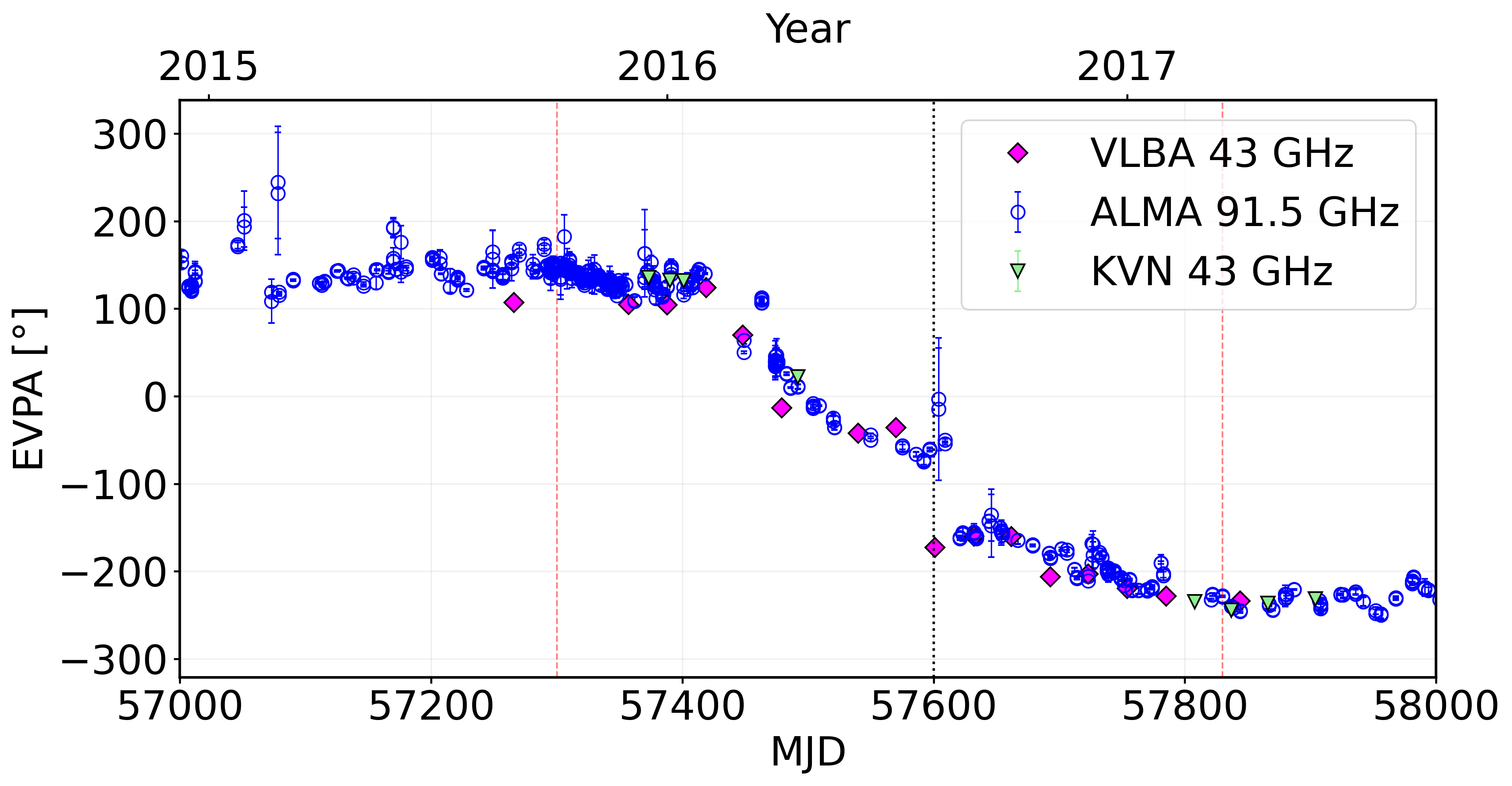}
\caption{Comparison of the core EVPA of OJ~287 measured with the BU 43~GHz VLBA monitoring program, the KVN at 43~GHz, and ALMA at 91.5~GHz. The BU measurements shown here span the interval before, during, and after the large EVPA rotation, allowing a direct comparison between the VLBI core EVPA and the contemporaneous KVN and ALMA measurements. The BU core EVPAs are broadly consistent with the KVN and ALMA EVPAs before and after the rotation and reproduce the systematic EVPA variation during the rotation.}
\label{app:fig_bu_evpa}
\end{figure*}

Figure~\ref{app:fig_bu_evpa} compares the BU 43~GHz core EVPA with the KVN 43~GHz and ALMA 91.5~GHz EVPAs. The BU core EVPAs are broadly consistent with the contemporaneous KVN and ALMA measurements both before and after the large rotation, and also reproduce the systematic EVPA variation during the rotation. This agreement indicates that the preferred EVPA orientations seen in the unresolved measurements are also present in the VLBI core.

\begin{figure*}[ht!]
\centering
\includegraphics[width=\textwidth]{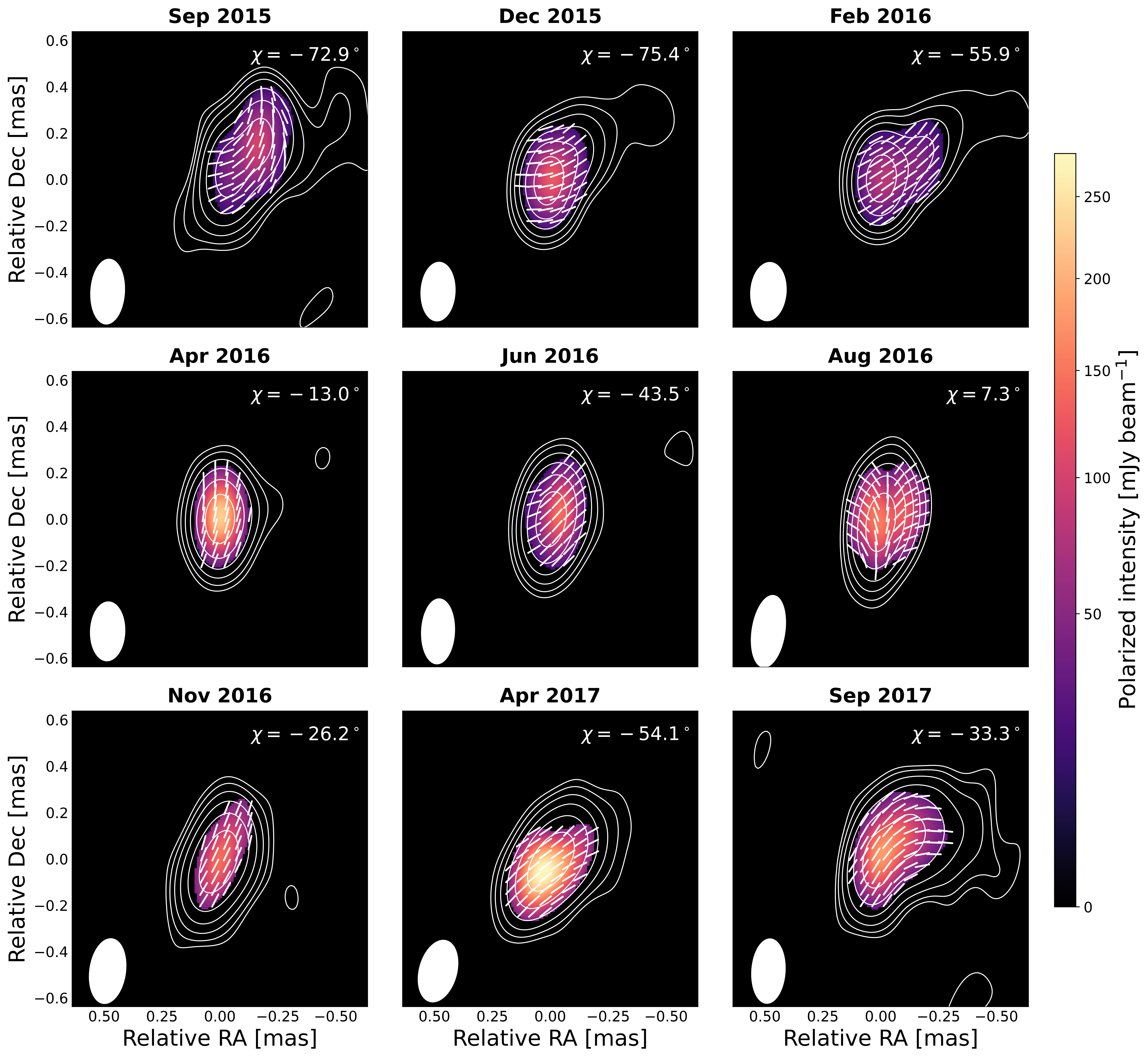}
\caption{Representative 43~GHz VLBA polarization images of OJ~287 at 9 epochs spanning the period before, during, and after the large EVPA swing. The color scale shows the polarized intensity (in log scale) in units of mJy~beam$^{-1}$, and the contours represent the Stokes $I$ intensity. 
The line segments indicate the local EVPA orientation. 
The restoring beam is shown in each panel, while $\chi$ indicates the 
measured core EVPA.}
\label{app:fig_bu_maps}
\end{figure*}

To illustrate the corresponding polarized structure of the source, we additionally show representative 43~GHz VLBA Stokes $I$ images with the EVPA distribution overlaid in Figure~\ref{app:fig_bu_maps}. Nine epochs were selected to sample the evolution before, during, and after the EVPA swing. The panels therefore provide a direct view of how the core polarization orientation evolves through the event.

\end{document}